\documentclass[referee]{raa}          

\usepackage{graphicx,times}             
\input{epsf.sty}                        
\input{psfig.sty}                       

  \newcommand{\vinf}{\mbox{$v_{\infty}$}}

  \newcommand{\kmsec}{\mbox{$\rm{km \; s^{-1}}$}}

  \newcommand{\msun}{\mbox{$M_\odot$}}

  \newcommand{\mdot}{\mbox{$\dot{M}$}}

  \newcommand{\Rstar}{\mbox{$R_*$}}

  \newcommand{\Rsun}{\mbox{$R_\odot$}}

  \newcommand{\sigT}{\mbox{$\sigma_{\rm T}$}}

  \newcommand{\ns}{\mbox{$\dot N $}}

  \newcommand{\nf}{\mbox{${\cal N}$}}

  \newcommand{\sr}{\mbox{${\cal R}$}}

  \def \etal{et~al.}

\begin{document}

  \title {Polarization Variability Arising from Clumps in the Winds of Wolf-Rayet  Stars}

   \volnopage{Vol.0 (200x) No.0, 000--000}      
   \setcounter{page}{1}  

\author{Q. Li
      \inst{1,3}\mailto{}
\and J.~P. Cassinelli
\inst{2}
\and J.~C. Brown
\inst{3}
\and R. Ignace
\inst{4}
}

\institute{Dept. of Astronomy, Beijing Normal University, Beijing 100875, China\\
             \email{qkli@bnu.edu.cn}
\and 
Dept. of Astronomy, University of Wisconsin-Madison, 53711, USA; cassinelli@astro.wisc.edu\\
\and
Dept. of Physics and Astronomy, University of Glasgow, Glasgow, G12 8QQ,
Scotland, UK; john@astro.gla.ac.uk\\
\and
Dept. of Physics and Astronomy, Eastern Tennessee State University, USA; ignace@etsu.edu\\
}

   \date{Received~~2001 month day; accepted~~2001~~month day}

 \abstract{ 
The polarimetric and photometric variability of Wolf-Rayet (WR) stars as caused by clumps in the winds, is revisited. In the model which is improved from Li \etal\ 2000, the radial expansion of the thickness is accounted for, but we retain the dependence on the $\beta$ velocity law, stellar occultation effects. We again search for parameters that can yield results consistent with observations in regards to the mean polarization $\bar p$, the ratio $\sr=\sigma_{\rm p}/\sigma_{\rm phot}$ of polarimetric to photometric variability, and the volume filling factor $f_V$. Clump generation and spatial distribution are randomized by the Monte Carlo method so as to produce clumps which are, in the mean, distributed uniformly in space and have time intervals with a Gaussian distribution. The generated clumps move radially outward with a velocity law determined by a $\beta$ index, and the angular size of the clumps is assumed to keep fixed. By fitting the observed $\sigma_{\rm p}/\sigma_{\rm {phot}}$ and the volume filling factor $f_V$, the clump velocity law index $\beta$ ($\sim 2$) and clump ejection rate $\nf$ ($\sim 1$) are inferred, and are found to be well constrained. In addition, the subpeak features on broad emission lines seem to support the clump ejection rate. Meanwhile, the fraction of the total mass loss rate that is contained in the clumps is obtained by fitting the observed polarization. We conclude that this picture for the clump properties produces a valuable diagnostic of WR wind structure.
}

   \authorrunning{Q. Li, J.P. Cassinelli, J.C. Brown \& R. Ignace}            
   \titlerunning{WR wind clump polarization variability}  

   \maketitle

\section{Introduction}
The intrinsic polarization of hot stars results from the scattering of starlight by electrons in aspherical stellar winds. There is evidence that the asymetry is not from a fixed structure but rather from stochastic ejection of clumps from the base of hot star winds. This evidence has been accumulating from a variety of spectropolarimetric 
studies that show that variability appears at a wide range of timescales of days, weeks, and months (Lupie \& Nordsieck 1987; Taylor \etal 1991). In the case of Wolf-Rayet stars, detailed observational data on the photometric, polarimetric, and spectral line profile variability have been presented and discussed by numerous authors (e.g., St.-Louis \etal 1987; Drissen \etal 1987;  Robert \etal 1989; Drissen \etal 1992; Moffat \& Robert 1992; Robert 1992; Lepine et al. 2000). Moffat \& Robert (1992) and Lepine \& Moffat (1999) have interpreted the Wolf-Rayet observations in terms of a distribution of dense clumps which result from hierarchical turbulence in the stellar winds. Brown et al. (1995) have discussed some of the physical properties of the larger clumps which dominate aspects of the data. Brown (1993) and Brown et al. (1999) conclude that clumps arise either by localized mass loss enhancements at the stellar surface or by the action of radiatively driven shocks sweeping up material on large scales. Brown et al. (2004) have discussed the combining effects of clumping and multiple scattering on the ``momentum paradox'' of WR star winds.

With the systematic monitoring campaigns for WR stars, Robert et al. (1989) and Robert (1992) find that statistically, WR stars display polarization about 0.1\%, and show broad band polarimetric variation ($\sigma_{\rm p}\approx 0.1\% - 0.02\%$) that is much smaller than the fractional photometric variability ($\sigma_{\rm phot}$), the mean ratio being about $\sr=\sigma_{\rm p}/ \sigma_{\rm phot}\approx 0.05$. Richardson et al. (1996) investigate the statistical effect of having larger numbers of clumps present and conclude that the clumps must be very dense so that their emission ($\propto n^2V$) is large enough to increase $\sigma_{\rm phot}$ and/or their optical depth is large enough to reduce $\sigma_{\rm p}$ by multiple scattering. Li et al. (2000) revisited the analysis by carrying out numerical simulations, following clump flow from the star with a $\beta$ velocity law while accounting for the occultation of clumps behind the stellar disk, but while also retaining the single scattering assumption. They concluded that the clumps follow a large $\beta$ velocity index, which means that the clumps are accelerating relatively slowly with radius in the wind. They also derived a certain range of clump ejection rate that along with $\beta$ allows for a fit to the statistical data. Valuable constraints on the WR parameters, such as mass loss rate, can be inferred from the observed mean polarization $\bar p$ (about 0.1\% level), the polarimetric and photometric variances. These conclusions are supported when considered in conjunction with the number of distinct narrow features that are seen in emission line profiles.

As for the observational evidence on clumping of hot stars in other wavebands, Ignace, Quigley, \& Cassinelli (2003) used the observations with the Infrared Space Observatory (ISO) SWS spectrometer to constrain the velocity law and wind clumping, and found that a $\beta$-value of the velocity law of the wind is about 2 -- 3 and the volume filling factor is up to 40\% for WR 136.  Marchenko et al. (2006) obtain numerous overdense clumps in the wind of WR 135 with FUSE observations. Chandra high resolution spectral observations of line profiles have shown the clumping wind properties of hot stars (Owocki \& Cohen 2006; Oskinova, Hamann, \& Feldmeier 2007; Cassinelli et al. 2008).

Recently, Davies et al. (2007) used the clump-ejection model to study the polarimetric variability of hot stars, in particular of luminous blue variables (LBVs), and got the detailed clump parameters through their simulations. One of their conclusions is that many tiny clumps are ejected around hot stars. Of particular interest to us is that Davies et al. (2007) find a flaw in assumptions of Li et al. (2000), in their assumption that both the thickness of a clump is fixed and that the number of electrons in a clump is constant. These results from Davies et al. (2007) motivate us to update the model of Li et al. (2000). It is our goal to understand the general nature of the Wolf-Rayet stars, using a picture for the ejection of clumps into the wind. Here we allow for the thickness of a clump to vary with the wind flow. We find that the main conclusions in Li et al. (2000) still hold, and that the present model leads to well constraints on the parameters, such as $\beta$ and $\nf$.

The basic model scenario is presented in Section 2, where the basic formulation concerned with the polarization and scattering light of both a single clump and ensemble of clumps are taken into account for various $\beta$ and $\nf$. In Section 3, the model results are presented and discussed, and in Section 4 general conclusions are presented.

\section {Clumping Model}

\subsection {Basic Formulation of Polarization for A Single Clump} 

\begin{figure}

\plotone{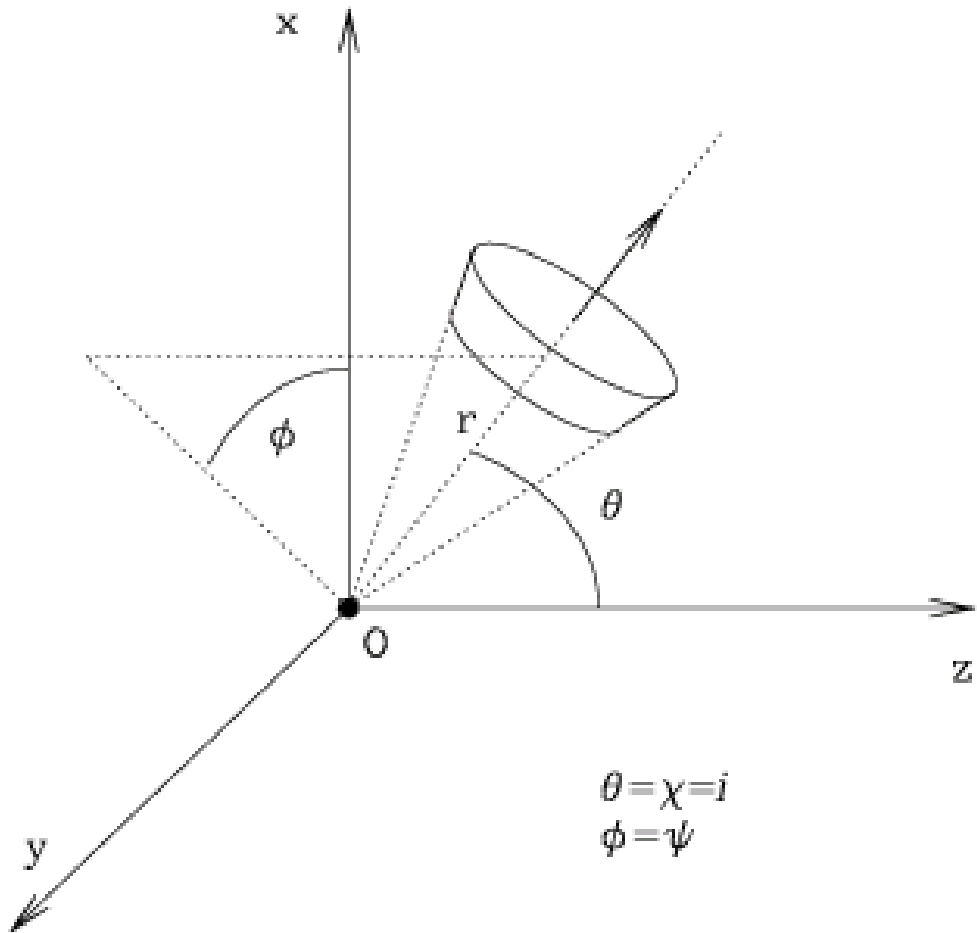}
\caption{ Geometry of scattering from one clump.  We employ a spherical coordinate system  $(r,\theta,\phi)$ with $oz$ along the line of sight. For each clump the
  ``inclination'' angle $i$ is identical to the scattering angle $\chi$
  as well as the polar angle $\theta$, while the polarization position angle
  on the sky, $\psi$, is the coordinate component $\phi$ (see Li et al. 2000).}
   \label{fig1}
\end{figure}

We assume that the polarimetric and photometric variability of a WR star is due to localized mass ejections at the stellar surface, over which the clumps are generated  at random positions, and at random time intervals with a normal distribution of mean value $\Delta t$. The clumps are then taken to move radially outward with a velocity law, thus the thickness $\Delta r$ of the clump should simultaneously expand along the radial wind flows, while the solid angle $\Delta \Omega$ is assumed constant (The geometry of a single outflowing clump is shown in Fig.~\ref{fig1}). The electron density then decreases as $r^{-2}\,v^{-1}(r)$. The clumps thus have axisymmetric shapes and, on the assumption that they are not optically thick in the continuum, the results of Brown \& Mclean (1977) can be used to find the polarization of a single clump as:

  \begin{equation}
  p = \tau_{\rm opt}\,(1-3\gamma)\sin^2 \it i,
\label{pbrown}
  \end{equation}
  where
  \begin{equation}
  \tau_{\rm opt}={3 \over 16} \sigT\, \int_{r_1}^{r_2}\int_{\mu_2}^{\mu_1}
  n(r,\mu)\, dr\,d\mu
  \end{equation}
is a mean optical depth, and

  \begin{equation}
  \gamma=\frac{\int_{r_1}^{r_2} \int_{\mu_2}^{\mu_1} n(r,\mu) \mu^2\,
  dr\,d\mu} {\int_{r_1}^{r_2}\int_{\mu_2}^{\mu_1} n(r,\mu)\, dr\,d\mu}
  \end{equation}
\noindent is a ``shape'' factor. $\it i$ is the clump axis inclination to the line of sight; $\mu=\cos\vartheta$, where $\vartheta$ is the clump opening angle between the axis of symmetry and the direction of the scatter seen from the center of the star; $n(r,\mu)$ is the electron number density in the clump; and $\sigT $ is the Thomson scattering cross-section. For the local reference frame ($r,\vartheta$) chosen, we let $\mu_1 = 1$.  Given constant solid angle $\Delta \Omega$, the electron density is  assumed to vary only with the distance, i.e., as $n(r,\mu) =n(r)$. To calculate the electron density in one clump, we use the mass conservation law:

  \begin{equation}
  \dot M_b=\Delta \Omega\,r^2\,\rho(r)\,v(r),
  \end{equation}
\noindent where $\dot M_b$ is the mass outflow rate  within one clump, $\rho(r)$ is the mass density, and $v(r)$ is the radial  velocity law that the clump will follow, which  we adopt to be of the common form:

\begin{equation}
  v(r)=\vinf\,\left(1-\frac{b\,\Rstar}{r}\right)^{\beta},
\end{equation}

\noindent where $\vinf$ is the terminal clump speed, $\Rstar$ is the photospheric radius of the WR star (We note that there are dynamical and effective optical photosphere notations, see Brown et al. 1995), $b$ is a dimensionless parameter to ensure that the initial wind speed is non-zero ($b=0.995$ is adopted throughout the simulations), and $\beta$ is a velocity law index, one of our basic clump parameters.

  The electron density in a clump thus becomes
  \begin {eqnarray}
  n_e &= &\frac {\rho} {\mu_e\,m_H} \nonumber \\
      &= & \frac {\dot M_b}{\mu_e\,m_H\,\Delta \Omega\,r^2\,v} \nonumber \\
      &= & \frac {\dot M_b}{\mu_e\,m_H\,\Delta \Omega\,r^2\,v},
\label{ne}
  \end {eqnarray}

where $m_H$ is the hydrogen mass, $\mu_e$ is the mean particle weight per free electron.

  To deal with the finite star geometry, the point source depolarization correction factors $D(r/R_\ast), C(r/R_\ast,\chi)$ can be employed according to Cassinelli et al. (1987) and Brown et al. (1989),

  \begin {eqnarray}
  D &=&\sqrt{1 - \frac {R_\ast ^2}{r^2}} \nonumber \\
  &=&\sqrt{1 - \frac {1} {x^2}}
  \end{eqnarray}
  and
  \begin {equation}
  C = \frac {8-D(1+D)\,(1- 3\cos^2\chi)}{3(1+D)(1+\cos^2\chi)},
\label {Ccorre}
  \end {equation}
where $x = r/ R_\ast$, again $R_\ast$ is the photospheric radius of the WR star, and $\chi$ is the scattering angle.

We wish to combine Eqs.~\ref{pbrown} to \ref{Ccorre} to yield an expression for the 
polarization  from a single clump with the assumed geometry. After the dimensionless 
treatment of $r$ (i.e., $r1/\Rstar=x1$ and $r2/\Rstar=x2$), the polarization expression now becomes

  \begin{eqnarray}
  p &=& \frac {3}{16}\,\sigT\,n_o\,\Rstar\,(1 - \mu_2)(\mu_2 +
  \mu_2^2) \sin^2\chi 
  \int_{x_1}^{x_2}\left(\frac {x} {x-b}\right )^{\beta}\,\frac {D(x)\,dx}{x^2}
     \nonumber \\
   &=&\frac {3}{16}\,\sigT\,n_o\, R_\ast\,(1 - \mu_2)(\mu_2 +
  \mu_2^2)  \sin^2\chi {}\nonumber\\
& &{}\times \int_{x_1}^{x_2} \left (\frac {x} {x-b}\right )^{\beta}{1 \over
  x^2}\,\sqrt{1-{1 \over
  x^2}}\,dx
 \end{eqnarray}
and the scattered light intensity $f_{\rm s}$ as a fraction of $L_\ast/4\pi$ in terms of Brown et al. (1995) is:

  \begin {eqnarray}
  f_{\rm s} &=&\frac {3}{16}\,\sigT\,n_o\,R_\ast\,(1 - \mu_2) (1+\cos^2\chi)
  \int_{x_1}^{x_2}\left(\frac {x} {x-b}\right )^{\beta}
  \frac {C(r,\chi)\,dx}{x^2}   \nonumber \\
      &=&\frac {3}{16}\,\sigT\,n_o\,R_\ast\,(1 -\mu_2) {}\nonumber\\
& &{} \times \int_{x_1}^{x_2}
  \frac {8-D(1+D)(1-3\cos^2\chi)}{3(1+D)x^2}\left (\frac {x}{x-b}\right)^{\beta}\,dx,
  \end{eqnarray}
where $n_o=\dot M_b /(\mu_e\,m_H\,\Delta \Omega\,\Rstar^2 \,\vinf)$.

\subsection{Inference of Conservation of Electrons in A Clump}

It is assumed that the clump has an initial extent of radial thickness $\Delta  x=x2-x1$  ($=0.01$, for example) and solid angle $\Delta \Omega$ ($=0.04$, for example), and the solid angle remains fixed. As the clump moves outward radially, obeying the velocity law, its thickness will naturally change to expand radially with both outer face and inner face obeying the local velocity expression. We may explicitly obtain the number of electrons in one clump

  \begin {eqnarray}
  N_e &=&\int_{0}^{2\pi}\int_{\mu1}^{\mu2}\int_{r1}^{r2} n_e dV  \nonumber\\
        &= & \frac {\dot M_b R_\ast}{\mu_e\,m_H\,\vinf} \int_{x1}^{x2}\frac{dx}{(1-b/x)^{\beta}}, 
\label {en}
  \end {eqnarray}
where $dV=-r^2\,d\mu\,d\phi dr$ in a local spherical coordinate, and for the geometry of the clump assumed, one may get $\mu1=1$ and $\Delta\Omega=2\pi(1-\mu2)$.

In Eq.~\ref{en}, for given values of $\beta$ and initial $\Delta x$, the integral can be solved analytically and/or numerically. In the work, we adopt three acceleration cases: $\beta=0.5$ for a rapidly accelerating flow; $\beta= 1$, a commonly assumed value for hot star winds; and $\beta=2$ for the slow acceleration case. These are chosen so we can determine which case provides an improved fit to the WR observational properties.

(1) In case of $\beta=0.5$, we may analytically do the integral in the Eq.~\ref{en} and obtain

  \begin {eqnarray}
 N_e &= & \frac {\dot M_b R_\ast}{\mu_e\,m_H\,\vinf} \int_{x1=1}^{x2}\frac{\sqrt{x}}{\sqrt{x-b}}dx\nonumber \\
 &= & \frac {\dot M_b R_\ast}{\mu_e\,m_H\,\vinf} \left[\sqrt{x(x-b)}+b\ln\left(\sqrt{x}+\sqrt{x-b}\right)+con1 \right]_{x1}^{x2}\nonumber \\
&= & N_{e0} \left(\sqrt{x2(x2-b)}-\sqrt{x1(x1-b)} 
+b\ln\frac{\sqrt{x2}+\sqrt{x2-b}}{\sqrt{x1}+\sqrt{x1-b}} \right)
 \nonumber \\
 &= & N_{e0} \left[\sqrt{x(x-b)}+b\ln\left(\sqrt{x}+\sqrt{x-b}\right)+con1 \right]_1^{1+0.01}, 
\label {en1}
\end {eqnarray}
where $con1$ is a constant from integral and $N_{e0}=\frac {\dot M_b R_\ast}{\mu_e\,m_H\,\vinf}$ is also a constant for given the stellar parameters. Owing to the boundary condition assumed, i.e. $x2=x1+\Delta x=1+0.01$ at $x1=1$, the expression 
$\left[\sqrt{x(x-b)}+b\ln\left(\sqrt{x}+\sqrt{x-b}\right)+con1 \right]_1^{1+0.01}$ equals to 0.103. Therefore, Eq.~\ref{en1} can clearly yield the following equation

\begin {equation}
\sqrt{x2(x2-b)}-\sqrt{x1(x1-b)}+b\ln\frac{\sqrt{x2}+\sqrt{x2-b}}{\sqrt{x1}+\sqrt{x1-b}}=0.103.
\label{betaha}
\end{equation}

(2) In case of $\beta=1$, we use the same method above and get the equation

\begin {equation}
x2-x1+b\ln\frac{x2-b}{x1-b}=1.103.
\label{beta1}
\end{equation}

(3) In case of $\beta=2$, we repeat the same process and obtain the following equation

\begin {equation}
x2-x1+2b\ln\frac{x2-b}{x1-b}+\frac{b^2(x2-x1)}{(x1-b)(x2-b)}=134.2.
\label {beta2}
\end{equation}

Eq.~\ref{betaha}, \ref{beta1}, or \ref{beta2} sets up the coherent relationship
between $x1$ and $x2$. We may use them with a given value of $\beta$ to get $x2$ 
once $x1$ is specified. In the simulations, we use the Newtonian bisection and bracketing methods to solve the Eq.~\ref{betaha}, \ref{beta1}, or \ref{beta2} due to the nonlinear relationship between $x1$ and $x2$. Note, it is clearly shown that the number of electrons in a clump with various $\beta$ are different ($N_e=N_{e0}\times 0.103$ for $\beta=0.5$, $N_e=N_{e0}\times 1.103$ for $\beta=1$ and $N_e=N_{e0}\times 134.2$ for $\beta=2$). [We may also have a special case of $\beta=0$, which will give the value of the integral equals to $\Delta x$ ($\Delta x$=0.01) and then we will have the expression $N_e=N_{e0}\times 0.01$ for $\beta=0$.] The electron scattering in the clump results in the polarization, hence the number of electrons will affect the strength of polarization, as discussed in Section 3.

\subsection {Relation between Time $t$ since Clump Expulsion and the Clump Distance $r$ from the Star}

The radial thickness varies with time as the clump moves out. We can connect the time since expulsion of a particular clump with its current radial distance together from the velocity law. Similarly, given time $t$, the clump distance (say $r$) can be obtained. From the clump velocity law, this relationship is,

  \begin {eqnarray}
 t &= & \int \frac{dr}{v(r)}\nonumber \\
 &= & \frac{R_\ast}{\vinf}\int \frac{dx}{(1-b/x)^{\beta}} \nonumber\\
&= & \tau\int \frac{dx}{(1-b/x)^{\beta}}, 
\label {rt}
  \end {eqnarray}
where $\tau=R_\ast/\vinf$ is denoted as the ``flow time scale''.

Again, in the case of $\beta=0.5$,

  \begin {eqnarray}
 \frac{t}{\tau} &= & \int \frac{\sqrt{x}}{\sqrt{x-b}}dx \nonumber \\
 &= & \sqrt{x(x-b)}-\frac{b}{2}\ln \left[\frac{x-b}{b}\left(\sqrt{\frac{x}{x-b}}-1\right)^2 \right]+con2.
\label {rtha1}
  \end {eqnarray}

If we adopt $x=1$ at $t=0$ and use $b=0.995$ so far, then we will get $con2=-0.14$ in the 
Eq.~\ref{rtha1}. Therefore, the expression between time and distance is

  \begin {equation}
 \frac{t}{\tau}= \sqrt{x(x-b)}-\frac{b}{2}\ln[\frac{x-b}{b}(\sqrt{\frac{x}{x-b}}-1)^2]-0.14.
\label {rtha}
  \end {equation}

In the cases of $\beta=1$ and $\beta=2$, the expression between time and distance are, respectively, 

  \begin {equation}
 \frac{t}{\tau} = x+b\ln(x-b)+4.27
\label {rt1}
  \end {equation}
and

  \begin {equation}
 \frac{t}{\tau} = x-b+2b\ln(x-b)-\frac{b^2}{x-b}+208.55.
\label{rt2}
  \end {equation}.

Therefore, from the numerical experiments, $t/\tau$ may be stepwise accumulated by doing summation of $\Delta t/\tau$, and $\Delta t/\tau$ is chosen to have a Gaussian distribution.

\subsection{Polarization for An Ensemble of Clumps}

The expressions above for polarization and scattering are for a single clump. For accounting for many clumps, we use the same approach as in Li et al. (2000). We set up a spherical coordinate system $(r,\theta,\phi)$ with the polar axis $oz$ being along the line of sight.  Then for each clump the ``inclination'' angle $\it i$ is identical to both the scattering angle $\chi$ and the polar angle  $\theta$, while the polarization position angle in the sky, $\psi$, is just the coordinate component $\phi$ (see Fig.~\ref{fig1}). 

For a system of clumps labeled $j=1,N$, the total scattered light fraction $f_{\rm s}$ and the net polarization are as usual for the optically thin case, simply given by the sum over $j$ the Stokes intensity parameters $Q_j=p_j\cos 2\phi_j$, $U_j=p_j\sin 2\phi_j$ of each to get the totals of $Q$ and $U$, then finding $p = (Q^2 + U^2)^{1/2}$ and position angle $\Psi ={1 \over 2} \arctan \frac {U}{Q}$. It is to be understood that {\it summations exclude all occulted clumps} -- i.e. clumps whose coordinates $x_j$, $\theta_j$, $\phi_j$ satisfy $\theta_j > \pi/2$ and $x_j\,\sin\theta_j < 1$.  Note that $\mu_j=\cos\theta_j$ is uniformly sampled in the interval $ -1$ to $+1$ and $\phi_j$ from 0 to $2\pi$ . The radii $x_j$ are determined by time and the velocity law.

The total mass loss rate $\mdot$ is distributed among all clumps and the ambient, inhomogeneous but, on average, spherical ``wind''. We choose the fraction of the mass loss in the clumps to be $\eta$. Hence,  $\eta \dot M = \ns N_e \mu_e m_H$, where $\ns$ is the mean clump ejection rate (s$^{-1}$) and $N_e$ again is the total number of electrons in one clump. Thus, if we fix $\dot M$ and increase $\ns$ then there are more clumps in any given range of $r$ but each would be of smaller $N_e$. We denote $\nf=\ns \,\tau$, the number of clumps ejected per characteristic flow time ($\tau =\Rstar/\vinf$), as the measure of the clump ejection rate. If we assume the portion of the total mass loss rate flowing into $k$ clumps and each of that has same number of electrons, then there is a scaling relationship $\eta \dot M =k \dot M_b$. Since in a flow time $\tau$, there are $\nf$ clumps ejected, we can attain $k\sim\nf$. Therefore, we may replace $\dot M_b$ in Eq.~\ref{ne} with $\eta \dot M/ \nf$ when accounting for many clumps in the simulations.

  The system of equations governing the time varying polarization and scattered light is then

  \begin {eqnarray}
  Q &=&\sum^{N}_{j=1}\,Q_j  \nonumber \\
  &   =& \frac {3}{16}\,\sigT\,n_o\, R_\ast(1 - \mu_2)(\mu_2 + \mu_2^2)
  \sum^{N}_{j=1}\sin^2\theta_j\,\cos 2\phi_j  {}\nonumber \\
& &{} \times \int_{x_1}^{x_2}\left (\frac
  {x_j} {x_j-b}\right )^{\beta}{1 \over x_j^2}\,\sqrt{1-{1 \over
  x_j^2}}\,dx_j  \nonumber  \\
  &   =& \frac {3}{16}\,\sigT \frac{n'_o}{\nf} R_\ast
(1 - \mu_2)(\mu_2 +  \mu_2^2)
  \sum^{N}_{j=1}\sin^2\theta_j\,\cos 2\phi_j  {}\nonumber \\
& &{} \times \int_{x_1}^{x_2}\left (\frac
  {x_j} {x_j-b}\right )^{\beta}{1 \over x_j^2}\,\sqrt{1-{1 \over
  x_j^2}}\,dx_j   \nonumber\\
&=&\frac{p_0}{\nf}\sum^{N}_{j=1}\sin^2\theta_j\,\cos 2\phi_j
  \int_{x_1}^{x_2}\left (\frac
  {x_j} {x_j-b}\right )^{\beta}{1 \over x_j^2}\,\sqrt{1-{1 \over
  x_j^2}}\,dx_j
  \end {eqnarray}

  and

  \begin {eqnarray}
  U& =&\sum^{N}_{j=1}\,U_j   \nonumber \\
  &   =& \frac {3}{16}\sigT\,n_o\, R_\ast(1 - \mu_2)(\mu_2 +  \mu_2^2)
  \sum^{N}_{j=1}\sin^2\theta_j\sin 2\phi_j  {}\nonumber\\
&  & {} \times \int_{x_1}^{x_2}\left (\frac
  {x_j} {x_j-b}\right )^{\beta}{1 \over x_j^2}\sqrt{1-{1 \over
  x_j^2}}\,dx_j       \nonumber   \\
  &   =& \frac {3}{16}\sigT \frac{n'_o}{\nf} R_\ast
(1 - \mu_2)(\mu_2 + \mu_2^2)
  \sum^{N}_{j=1}\sin^2\theta_j\sin 2\phi_j  {}\nonumber\\
&  & {} \times \int_{x_1}^{x_2}\left (\frac
  {x_j} {x_j-b}\right )^{\beta}{1 \over x_j^2}\sqrt{1-{1 \over
  x_j^2}}\,dx_j   \nonumber\\
&=& \frac{p_0}{\nf}\sum^{N}_{j=1}\sin^2\theta_j\sin 2\phi_j 
  \int_{x_1}^{x_2}\left (\frac
  {x_j} {x_j-b}\right )^{\beta}{1 \over x_j^2}\sqrt{1-{1 \over
  x_j^2}}\,dx_j,
  \end {eqnarray}

where $n'_o=\eta \dot M /(\mu_e\,m_H\,\Delta \Omega\,\Rstar^2 \,\vinf)$ and

\begin{eqnarray}
p_0&=&\frac {3}{16}\sigT n'_o R_\ast(1 - \mu_2)(\mu_2 + \mu_2^2) \nonumber\\
&=&\frac {3}{16}\sigT \frac{\eta \dot M}{\mu_e\,m_H\,\Delta \Omega\Rstar^2 \vinf} R_\ast(1 - \mu_2)(\mu_2 + \mu_2^2).
\label{p0}
\end {eqnarray}

Hence, the total polarization is given

  \begin {equation}
  p = \sqrt {Q^2+U^2}
\label{pol}
  \end {equation}

  and

  \begin {eqnarray}
  f_{\rm s}&=&\sum^{N}_{j=1}\,f_{\rm {sj}}  \nonumber \\
  &=&\frac {3}{16}\sigT n_o R_\ast(1 -
  \mu_2) {}\nonumber \\
& &{}\times \sum^{N}_{j=1}\int_{x_1}^{x_2}
  \frac {8-D_j(1+D_j)(1-3\cos^2\theta_j)}{3(1+D_j)x_j^2}\left (\frac
  {x_j}{x_j-b}\right )^{\beta}dx_j  \nonumber \\
  &=&\frac {3}{16}\sigT \frac{n'_o}{\nf} R_\ast (1 -
  \mu_2)  {}\nonumber \\
& &{}\times \sum^{N}_{j=1} \int_{x_1}^{x_2}
  \frac {8-D_j(1+D_j)(1-3\cos^2\theta_j)}{3(1+D_j)x_j^2}\left (\frac
  {x_j}{x_j-b}\right )^{\beta}dx_j        \nonumber\\
&=&\frac{f_0}{\nf}\sum^{N}_{j=1} \int_{x_1}^{x_2}
  \frac {8-D_j(1+D_j)(1-3\cos^2\theta_j)}{3(1+D_j)x_j^2}\left (\frac
  {x_j}{x_j-b}\right )^{\beta}dx_j,
\label{pho}
  \end{eqnarray}

where 

  \begin{eqnarray}
f_0&=&\frac {3}{16}\sigT n'_o R_\ast (1 -\mu_2) \nonumber\\
&=&\frac {3}{16}\sigT \frac{\eta \dot M}{\mu_e\,m_H\,\Delta \Omega\,\Rstar^2 \,\vinf} R_\ast (1 -\mu_2).
\label{f0}
  \end{eqnarray}

Note again, in the above expressions, both $x1$ and $x2$ are varying with time, so we need to use Eq.~\ref{betaha}, \ref{beta1}, or \ref{beta2} in regards to various $\beta$ to determine $x2$ once $x1$ is given. Here, $x1$ is obtained from Eq.~\ref{rtha}, \ref{rt1}, or \ref{rt2} in case of different $\beta$, if time $t$ is known from the numerical experiments.

On inspection of the above equations for polarization and scattered light, we expect that, for given $\dot M$, $\Delta \Omega$, initial $\Delta r$, and $\beta$, results should depend mainly on $\cal N$. For a fixed mass loss rate and flow time, if the clump generation rate is low, only a few clumps each of large density will be present near the star and these dominate the $p$ and $f_{\rm s}$ values, as is shown in Li et al. (2000). But for high generation rates, many low density clumps near the star will be controlling $p$ and $f_{\rm s}$. So the same total number of electrons is redistributed in different number of clumps, resulting in different statistical means and variances in the polarization and scattered light fraction. For a fixed clump ejection and mass loss rate, the number of clumps in the inner radii near the star has a steady mean value and so therefore do the resulting polarization, scattered intensity and their variances, but these values change with  $\cal{N}$, $\dot M$, and $\beta$. So their observed values allow inference of the clump emission and flow parameters.

\section {Model Results and Discussion}

\begin{figure}
\plotone{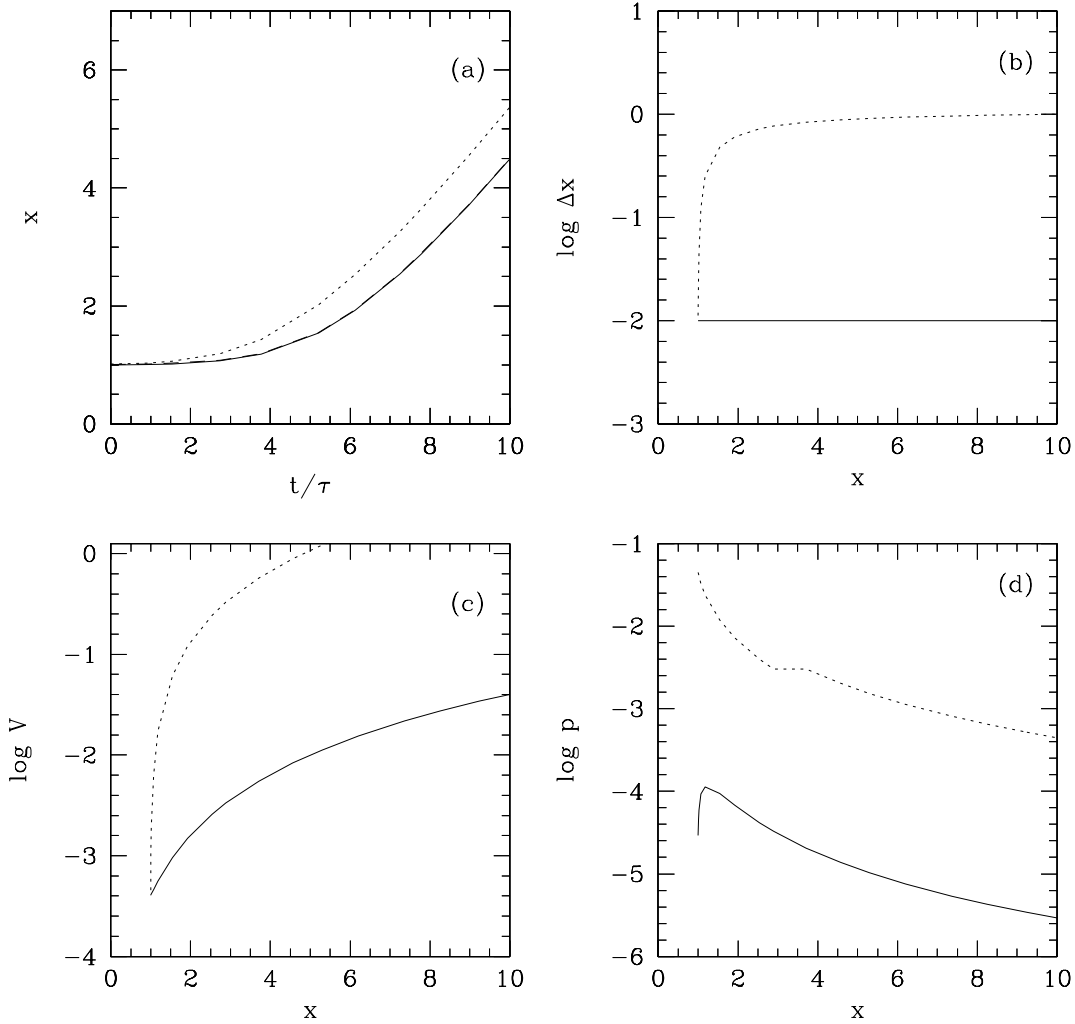}
\caption{Model results are shown for a single clump: (a) The radial extent vs time. 
The solid line denotes the inner radial boundary (say $x1$) and the dotted line denotes the outer radial boundary,  $x2$ ($x2=x1+\Delta x$) with the 
constant thickness $\Delta x=0.01$, is shown in the dashed line which is almost overlap the solid line $x1$. (b) The clump thickness versus the clump location. The solid line denotes the constant clump thickness as assumed. In contrast, the dotted line shows the expansion of the thickness with its location (i.e. $x$). (c) The clump volume vs its location. The solid line denotes the volume of the clump with constant thickness $\Delta x=0.01$ and the dotted line denotes the volume of the clump with expanding thickness. (d) Polarization vs the clump location. The dotted line is for the varying thickness and the solid line is for the constant thickness as in Davies et al (2007). }
   \label{fig2}
\end{figure}

\begin{enumerate}

\item
As a start, let us calculate the thickness, volume and polarization of a single clump with varying clump location, supposing $\sin\chi=1$ and $\beta=1$. The results are 
displayed in Fig~\ref{fig2}. In Fig.2a the solid line denotes the inner radial boundary (say $x1$) and the dotted line denotes the outer radial boundary (say $x2$).
We see that as the time progresses, $x2$ increases faster than $x1$, which directly 
induces the thickness expansion. In Fig.2a, as for comparison, $x2$ ($x2=x1+\Delta x$)
with the constant thickness $\Delta x=0.01$ is shown in the dashed line which almost overlaps the solid line $x1$. In Fig.2b, the solid line denotes the case for which a constant clump thickness is assumed. In contrast, the dotted line shows the expansion of the thickness with its location (i.e. $x$), and we see that the thickness increases rapidly among $x=1$ to $x=2$, where the winds are mainly accelerated. In Fig.2c, the solid line denotes the volume of the clump with constant thickness $\Delta x=0.01$ and the dotted line denotes the volume of the clump with expanding thickness, which is increasing dramatically faster than the constant thickness case.  We compare the polarization of varying thickness (the dotted line in Fig.2d) with that in the Davies et al. (the solid line in Fig.2d) for one single clump and see that in both cases, the polarization decreases as the clump moves outward, while there is a peak polarization about $x=1.2$ for the latter case.

\item
Furthermore, we calculate the thickness, volume and polarization of a single clump with varying clump location, supposing $\sin\chi=1$ but with different $\beta$, as is shown in Fig.3. To see how the $x1$ varies with time (say $t/\tau$) in case of differing $\beta$s, we show the results in Fig.3a in which fast acceleration of $\beta=0.5$ is displayed in dashed line and moderate acceleration of $\beta=1$ is in solid line, and slow acceleration of $\beta=2$ is in dash-dotted line. We also show the radial thickness varies with location of the clump in Fig.3b. The dash-dotted line denotes the case of $\beta=2$, the solid line denotes the case of $\beta=1$, and the dashed line denotes the case of $\beta=0.5$. In addition, we show the constant thickness case with $\beta=1$ in dotted line. In Fig.3c, we show that the volume changes with the clump location in various cases of $\beta$ as denoted in the figure. In Fig.3d, we show the polarization varies with location of the clump. The dash-dotted line denotes the case of $\beta=2$, the solid line denotes the case of $\beta=1$, and the dashed line denotes the case of $\beta=0.5$. In addition, we show the constant thickness case in dotted line. Note that the start values of polarization in various case of $\beta$ are different since the electron number in one single clump for various case of $\beta$ is different, as is shown from Eq.~\ref{en1} which is related to $\beta$. And polarization is mainly determined by the number of electrons in the clump (Brown et al. 1995).

\item
Taking into account the behavior of an ensemble of clumps, we compute the polarization, the fraction of scattering light intensity, and their variation. In particular, we calculate the ratio of their variations for various clump ejection rate $\nf$ in a flow time with various $\beta=0.5,\, 1,\, 2$. Interestingly, $\sigma_{\rm p}/\bar p$ and $\sigma_{\rm p}/\sigma_{\rm {phot}}$ do not depend on any specific star, but only on the $\nf$ and $\beta$. It is not surprising at this since $p$ (or $f_s$) linearly relies on $p_0$ (or $f_0$) which is determined by the stellar parameters of a star. So we enable to treat $\sigma_{\rm p}/\sigma_{\rm {phot}}$ as a probe to explore the probable $\nf$ and $\beta$. To get rid of the influence
of any specific star, we divide $p$ (as well as Q and U) in Eq.~\ref{pol} by $p_0$ in Eq.~\ref{p0} and divide $f_s$ in Eq.~\ref{pho} by $f_0$ in Eq.~\ref{f0}. The results are displayed in Tables~\ref{tab1}, \ref{tab2} and \ref{tab3}. Using the values in these tables, we plot $\sigma_{\rm p}/\bar p$ and $\sigma_{\rm p}/\sigma_{\rm {phot}}$ with various $\nf$ in cases of various $\beta=0.5,\, 1,\, 2$. The observed value of $\sigma_{\rm p}/\bar p = 0.5$ is denoted in dotted line in the upper panel of Fig.~\ref {fig4}, and the observed value of $\sigma_{\rm p}/\sigma_{\rm {phot}}=0.05$ is denoted in dotted line in the lower panel of Fig.~\ref {fig4}. We found that to achieve the observed value, three patterns could hold, for $\beta=0.5 \longrightarrow \nf\sim 500$; for $\beta=1 \longrightarrow \nf\sim 150$; and for $\beta=2 \longrightarrow \nf \sim 1$. Although there seems no solid unique solution for $\beta$ and $\nf$ in terms of the above considerations, while combining the effects of $\nf$ and $\beta$ on $\sigma_{\rm p}/\sigma_{\rm {phot}}$ and $f_V$ the volume filling factor, it seems the observed data might prefer the case of $\nf=1$ and $\beta=2$. We will further demonstrate this later.

Hamann and Koesterke (1998) found the clump volume filling factor of WR subtype WN stars about 30\% in terms of their spectrum analyses. In the work of the paper, we are able to calculate the clump volume filling factor from our model. After doing summations of volumes that all clumps occupy and divided it by the whole space accounted for, then the volume filling factor denoted as $f_V$, can in common definition be obtained,

\begin{equation}
f_V=\frac{\sum_{i=1}^{N}\Delta V_i}{\frac{4\pi}{3}(r^3-R_*^3)}.
\label{fv}
\end{equation}

We present the model results of $f_V$ in column 9 of Tables~\ref{tab1}, \ref{tab2} and \ref{tab3} in case of various $\beta$s. We also plot $f_V$ vs $\nf$ in Fig.~\ref{fig5}, in which the dotted line denotes the ``observed" of 30\%. The figure shows $\beta=0.5 \longrightarrow \nf \sim 700$; $\beta=1 \longrightarrow \nf \sim 100$; $\beta=2 \longrightarrow \nf \sim 1$, in regards to the observed. Note, $f_V$ greater than 1 means clumps overlapped or merged. So for the assumption of thin clumps, we would rule out the cases of $f_V>1$.

In general, the hot star winds show two-component property: the smooth ambient and clumps. Of course, when the clumps move outward to very far away distance, they will become as part of the interstellar medium. However, the density of clumps is in fact higher than that of the ambient wind around the star within several hundred of stellar radius. The smaller the value of the volume filling factor, the stronger the wind clumping. A clumping wind may cause higher emission for the same amount of material, which implies that the mass loss rates by spectroscopic analysis under the assumption of a smooth wind are systematically overestimated by typically a factor of $\frac{1}{\sqrt{f_V}} \sim 2$ and even higher, and the mass loss rates in turn consequently affect the stellar structure and evolution. Such clumping presumably arises from the inherent instability of radiation driven winds and would influence the strength of the electron scattering wings, change ionization and line ratios, and cause polarization variability and profile variability. In fact, Abbott et al. (1981) has taken the inhomogeneities (i.e. clumps) in the wind into account and got the relationship of the mass loss rate, radio flux, and the filling factor. Later, in most of models, in most simplicity, the inter-clump medium is usually treated as void, which recently leads to the porous scenario (Owocki and Cohen, 2006) for OB stars, as the photons are able to leak freely through the large separation between clumps. However, our model clearly indicates that the void treatment for WR stars might likely cast some doubts, since the inter-clump medium, occupying a very large mass percentage of the total, may play a role in the hot wind emission, in particular, near the star. The whole inter-clump medium, being of the smooth spherical wind ambient, globally contributes nil to the net polarization due to the cancellation effect of polarization.

\item 
We compute the photometric and polarimetric intensity and their fluctuations in case of various $\beta$s. Figs.~\ref{fig6}, \ref{fig8} and \ref{fig10} show how  polarization, position angle, and scattered light intensity change with the total number of clumps emitted from the start. Polarization changes are also shown as a locus in the Q -- U plane (see Figs.~\ref{fig6}, \ref{fig8} and \ref{fig10}) which, as expected, shows no strong preferred direction, since the mean structure is quasi-spherical.
In Figs.~\ref {fig7}, \ref {fig9}, and \ref {fig11} we show ``observational'' time--smoothed  results for the variations in mean polarization and scattered light. 
Standard deviations, $\sigma$, of these quantities are also plotted in these figures. 
The ratio $\sr$ versus total number of clumps $N$ as time progresses is plotted in Fig.~\ref{fig12}, from which we see that in order to obtain the $sustainable$ $\sr=0.05$ as observed, the pattern of $\beta=2$ and $\nf=1$ is indeed preferred. This seems comparable to the number of subpeaks of the broad line observed. Given the model quantities $\Rstar=10\Rsun$ and $\vinf=1800\,\kmsec$, the flow time $\tau$ is about 4000 seconds. If in one flow time, there is one new clump occurring ($\nf \sim 1$), then in 10 flow time, 10 new clumps will occur. Hence, there would consequently result in 10 more detectable subpeaks since the clump emission is dominated by the inner clumps in the line emission regions (LER), which are closer to the star. This conclusion is indeed consistent with the observations and their analysis (Robert 1992, Brown et al. 1995). Note, envisaging Tables~\ref{tab1}, \ref{tab2} and \ref{tab3} in which $\sr=0.05$ seem to be attained, however, observing Figs.~\ref {fig6}, \ref {fig8}, \ref {fig10}, and their time averaged results in Figs.~\ref {fig7}, \ref {fig9}, \ref {fig11}, in particular \ref {fig12}, we found that $only$ the case of $\beta=2$ and $\nf=1$ shows a real and sustainable result. Neither the case of $\beta=1$ and $\nf=200$, nor the case $\beta=0.5$ and $\nf=1000$, could achieve a long-standing $\sr=0.05$. Therefore, we have to abandon these cases. If we plot the number of clumps versus the distance (Fig.~\ref{fig13}), we will find that in case of various $\beta=0.5,\,1,\, 2$ but with same $\nf=1$, there will several hundred of clumps radially staggering outward from $1\Rstar$ to $5\Rstar$ in case of $\beta=2$ being very slowly accelerated, but just a few in case of $\beta=0.5,\,1$ being rapidly accelerated. In terms of the concept of LER in Lepine \&  Moffat (1999) and Dessart \& Owocki (2005), the LER with velocity space is about among 0.4--0.9$\vinf$ where the clumps are accelerated. We note that the WR-wind acceleration length scale $\beta \Rstar \sim 20\Rsun$ in their results, is compatible with our results, with $\Rstar=10\Rsun$ applied and $\beta=2$ inferred.

\item
For the fraction factor of the mass loss rate into the clumps $\eta$ in regards to $p_0$ in Eq.~\ref{p0} and $f_0$ in Eq.~\ref{f0}, we could use the typical polarization observed for WR stars $\bar p=0.1 - 1\%$ to constrain it. If the following typical parameters are employed: $\dot M=2.5\times 10^{-6} \msun$/year, $\Delta \Omega=0.04$, initial $\Delta r_0=0.01\Rstar$, $\mu_e=2$, $\Rstar=10\Rsun$ and $\vinf=1800\,\kmsec $, then we may gain $p_o=0.0147\eta$. From the above discussion, the pattern $\beta=2$ and $\nf=1$ is preferred, then we check Table~\ref{tab3} and get $\bar p'=\bar p/p_0\sim 124.85\longrightarrow \bar p=1.84\eta$. Therefore, we enable to gain $\eta=\bar p/1.84\sim 10^{-3} -- 10^{-2}$. This would imply that only a quite small portion of the winds is going to deposit into the clumps but most of winds are blown as an ambient where clumps embedded. In contrary, one might recall the solar wind picture in which the solar wind is a coherent outward expansion of the solar corona, frequently involving the corona mass ejection (CME) events. Hence, it is not surprising that there are two components in the hot star environments, being the ambient winds and the clumps (or wind-blown bubbles). However, the formation mechanisms of the solar winds and hot star winds are distinctly different. The former is driven by the gas pressure gradient of the high temperature solar corona, and the latter is driven by the pressure of the radiation emitted by the hot star, so-called the continuum-driven and line-driven. Note that the physical quantities in Tables~\ref{tab1}, \ref{tab2} and \ref{tab3}, such as $\bar p'=\bar p/p_0$, $\sigma_{\rm p}'=\sigma_{\rm p}/p_0$, $\bar f_{\rm s}'=\bar f_{\rm s}/f_0$, $\sigma_{\rm phot}'=\sigma_{\rm phot}/f_0$, $\bar p/\bar {f_s}$, $\sigma_{\rm p}/\sigma_{\rm phot}$, scale with $p_0$ or $f_0$, so they are dimensionless numbers. The fraction factor $\eta$ only affects the values of $p_0$ and $f_0$. Thus for a specific star one could gain $\eta$ incorporating the stellar parameters and observed polarization. We realize that the clumps ejected in our model are massive- or macro-clumps, distinguished from the local perturbations in the winds. 

\end{enumerate}

  \section {Conclusions}

In this paper, we update the previous model proposed by Li et al. (2000) and apply the clump ejection scenario to explain the WR star wind polarization and its variability observed, by accounting for the expansion of the clumps along the wind flows but keeping the solid angle constant. We may gain the main conclusions as follows.

(1) From numerous model simulations using various $\beta$ and $\nf$, incorporating with the volume filling factor $f_V$, we found $\beta\sim 2$ and $\nf\sim 1$ are preferred for explaining the observational data on photometric, polarimetric, and spectral line profile variability of WR stars. This gives well model constraints on the stellar wind properties of hot stars.

(2) We also found that a small fraction as $10^{-3}$ of the wind material deposits into the clumps but most of the mass loss are going to the space as wind ambient. This quantitative estimation of the mass fraction into the wind ambient implies that the porous wind models for WR stars might be cautious since the inter-clump medium is far from void. 

In summary, this updated model which is improved from Li \etal\ 2000, with inclusion of the radial expansion of the thickness of clumps but retaining the dependence on the $\beta$ velocity law and stellar occultation effects, yields results consistent with observations in regards to the mean polarization $\bar p$, the ratio $\sr=\sigma_{\rm p}/\sigma_{\rm phot}$ of polarimetric to photometric variability, and the volume filling factor $f_V$. It offers a quantitative estimation of the mass fraction into the clumps and ambient winds as well. Hence, the model produces a valuable diagnostic of WR wind structure.

\begin{acknowledgements}
We would like to thank the anonymous referee for the constructive comments that led 
to a significant improvement in the paper. The authors would like to thank Ben Davies for informative discussion for the work. The authors acknowledge support for this work from: the Natural Science Foundation of China grants 10273002, 10573022, and 10778601 (QL); the NSF Center for Magnetic Self Organization in Laboratory and Astrophysics Plasmas (JPC); UK STFC Rolling Grant (JCB). JPC and RI have been supported in part by award TM3-4001 issued by the $Chandra$ X-ray Observatory Center.
\end{acknowledgements}


  \begin{table}

  \begin{center}

  \caption[]{Simulation results for finite star source with occultation and velocity law with $\beta=0.5$ \label{tab1}}
  \vspace{1ex}
  \begin {tabular} {cccccccccc}
  \hline\hline
  N&$\nf$&$\bar p'$&$\sigma_{\rm p}'$&$\bar
  f_{\rm s}'$&$\sigma_{\rm phot}'$
  &$\overline p/\overline {f_{\rm s}}$
  &$\sigma_{\rm p}/\sigma_{\rm phot}$&$f_V$\\
  \hline

5000 &1/4 &0.8394 &0.8632 &4.928 &4.470 &0.337 &0.382 &0.0001\\

5000 &1 &0.3565 &0.2032 &2.319 &1.138 &0.305 &0.353 &0.0003\\

5000 &5 &0.1557 &0.0816 &1.927 &0.276 &0.160 &0.479 &0.0016\\

5000 &10 &0.1091 &0.0557 &1.891 &0.234 &0.114 &0.473 &0.0033\\

5000 &50 &0.0500 &0.0265 &1.835 &0.195 &0.054 &0.271 &0.0164\\

5000 &100 &0.0364 &0.0204 &1.783 &0.239 &0.040 &0.168 &0.0328\\

5000 &200 &0.0268 &0.0146 &1.693 &0.304 &0.031 &0.094 &0.0654\\

5000 &1000 &0.0143 &0.0051 &1.229 &0.424 &0.023 &0.024 &0.3190\\

5000 &2000 &0.0102 &0.0031 &0.910 &0.392 &0.022 &0.016 &0.6239\\

  \hline\hline
  \end{tabular}
  \end{center}
  Notes: Column 1 is the total number of clumps applied in the simulations. Column 2 is the clump ejection rate in a flow time. Column 3 is $\bar p'=\bar p/p_0$. Column 4 is 
$\sigma_{\rm p}'=\sigma_{\rm p}/p_0$. 
Column 5 is $\bar f_{\rm s}'=\bar f_{\rm s}/f_0$. Column 6 is $\sigma_{\rm phot}'=\sigma_{\rm phot}/f_0$. Column 7 is $\bar p/\bar {f_s}$. Column 8 is $\sigma_{\rm p}/\sigma_{\rm phot}$. Column 9 is the volume filling factor $f_V$.
  \end{table}

  \begin{table}
  \begin{center}
  \caption[]{Simulation results for finite star source with occultation and velocity law with $\beta=1$ \label{tab2}}
  \vspace{1ex}
  \begin {tabular} {cccccccccc}
  \hline\hline
  N&$\nf$&$\bar p'$&$\sigma_{\rm p}'$&$\bar
  f_{\rm s}'$&$\sigma_{\rm phot}'$
  &$\overline p/\overline {f_{\rm s}}$
  &$\sigma_{\rm p}/\sigma_{\rm phot}$&$f_V$\\
  \hline

5000 &1/4 &2.8922 &2.1003 &22.762 &17.995 &0.252 &0.231 &0.0015\\

5000 &1 &1.4512 &0.7755 &15.340 &5.933 &0.187 &0.259 &0.0035\\

5000 &5 &0.6270 &0.3205 &14.144 &2.371 &0.088 &0.268 &0.0175\\

5000 &10 &0.4432 &0.2223 &14.077 &1.753 &0.062 &0.251 &0.0350\\

5000 &50 &0.2066 &0.1172 &13.805 &1.664 &0.030 &0.140 &0.1725\\

5000 &80 &0.1641 &0.0982 &13.565 &1.985 &0.024 &0.098 &0.2727\\

5000 &100 &0.1458 &0.0904 &13.395 &2.170 &0.022 &0.082 &0.3381\\

5000 &150 &0.1189 &0.0703 &13.021 &2.557 &0.018 &0.054 &0.4962\\

5000 &200 &0.1054 &0.0510 &12.663 &2.852 &0.016 &0.035 &0.6460\\

5000 &400 &0.0788 &0.0279 &11.316 &3.529 &0.014 &0.016 &1.1568\\

5000 &1000 &0.0506 &0.0153 &8.065 &3.647 &0.013 &0.008 &2.2168\\

5000 &2000 &0.0255 &0.0088 &4.986 &2.591 &0.010 &0.007 &4.0601\\

  \hline\hline
  \end{tabular}
  \end{center}
  Notes: Parameters have the same meaning as that in Table~\ref{tab1}.
  \end{table}

  \begin{table}
  \begin{center}
  \caption[]{Simulation results for finite star source with occultation and velocity law with $\beta=2$ \label{tab3}}
  \vspace{1ex}
  \begin {tabular} {cccccccccc}
  \hline\hline
  N&$\nf$&$\bar p'$&$\sigma_{\rm p}'$&$\bar
  f_{\rm s}'$&$\sigma_{\rm phot}'$
  &$\overline p/\overline {f_{\rm s}}$
  &$\sigma_{\rm p}/\sigma_{\rm phot}$&$f_V$\\
  \hline

5000 &1/4 &247.339 &126.392 &17517.453 &3157.450 &0.028 &0.079 &0.1071\\

5000 &1 &124.8503 &64.1338 &17300.174 &1918.061 &0.014 &0.066 &0.4098\\

5000 &5 &48.1358 &28.5260 &16625.262 &2933.660 &0.006 &0.019 &1.7094\\

5000 &10 &33.3430 &17.1025 &15757.138 &3900.831 &0.004 &0.009 &2.6139\\

5000 &50 &14.3191 &3.7360 &9381.283 &4799.212 &0.003 &0.002 &1.3993\\

5000 &80 &9.6549 &2.6050 &6372.940 &3492.983 &0.003 &0.002 &3.3746\\

5000 &100 &7.8479 &2.1951 &5201.766 &2891.146 &0.003 &0.002 &5.6845\\

5000 &150 &5.2524 &1.4971 &3550.873 &2009.073 &0.003 &0.002 &12.349\\

5000 &200 &3.8611 &1.0675 &2689.970 &1536.053 &0.003 &0.001 &11.362\\

5000 &400 &1.9001 &0.5064 &1352.113 &778.007 &0.003 &0.001 &15.412\\

5000 &1000 &0.7599 &0.2026 &540.961 &311.203 &0.008 &0.001 &15.406\\

5000 &2000 &0.3800 &0.1013 &270.565 &155.602 &0.003 &0.001 &15.396\\

  \hline\hline
  \end{tabular}
  \end{center}
  Notes: Parameters have the same meaning as that in Table~\ref{tab1}.
  \end{table}

  \newpage

   \begin{figure}
\plotone{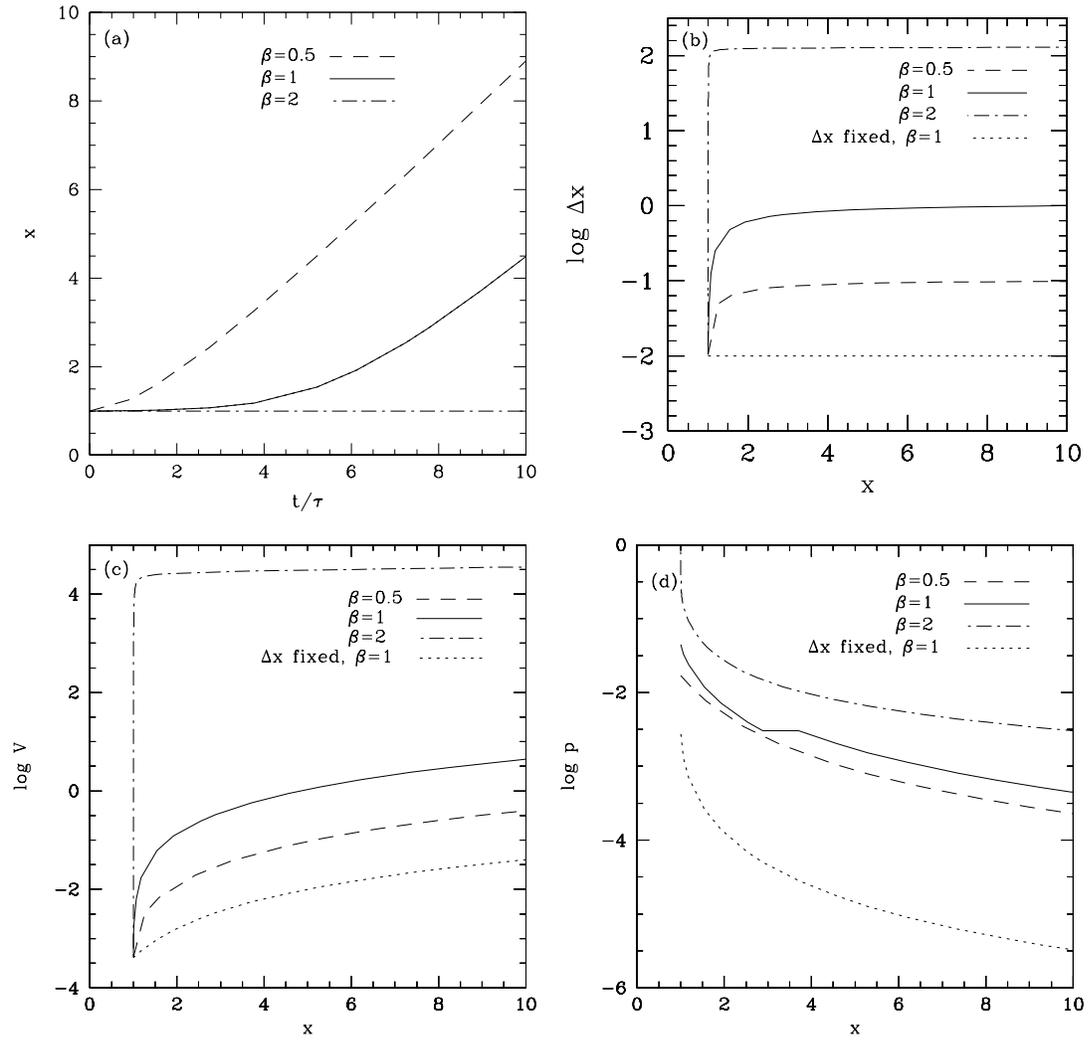}
\caption{Model results are displayed in various cases of $\beta$ as denoted in the figure for one single clump.}
   \label{fig3}
\end{figure}

   \begin{figure}
\plotone{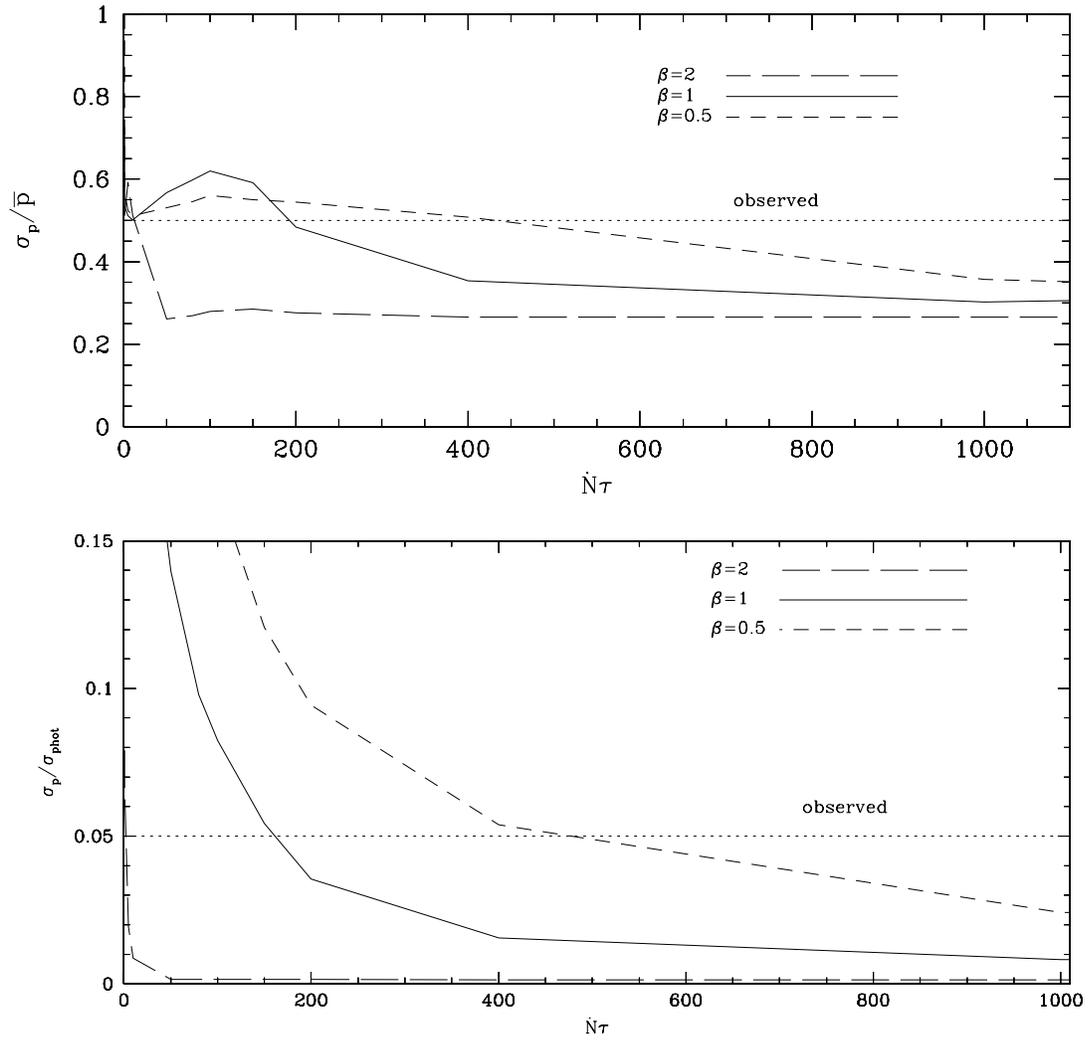}
\caption{The upper panel is $\sigma_{\rm p}/\bar p$ vs $\nf$ in various cases of $\beta$ as denoted in the plot. The lower panel is $\sigma_{\rm p}/\sigma_{\rm {phot}}$ vs $\nf$. The observed are denoted in dotted lines in two panels.}
   \label{fig4}
\end{figure}

   \begin{figure}

\plotone{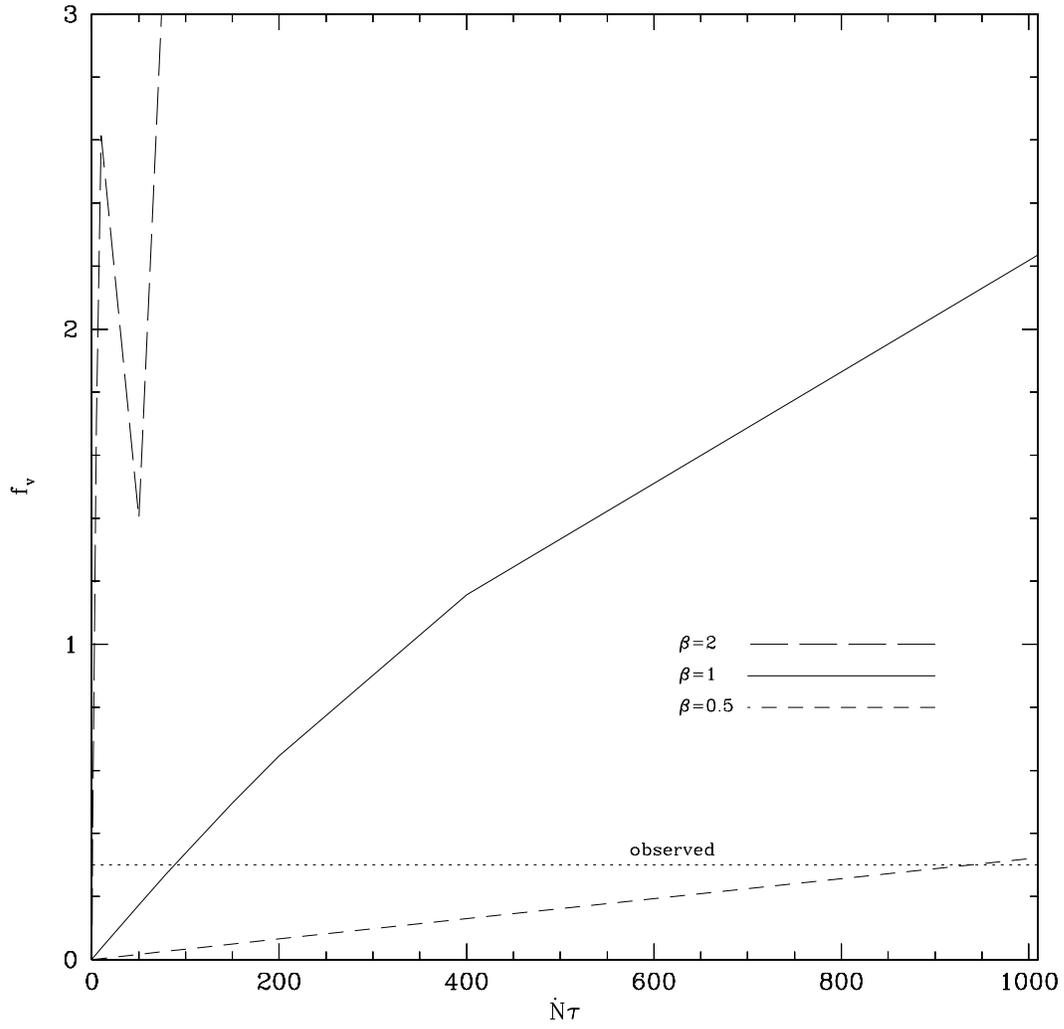}
\caption{Volume filling factor $f_{\rm v}$ vs $\nf$ in various cases of $\beta$ as denoted in the plot.}
   \label{fig5}
\end{figure}

   \begin{figure}
\plotone{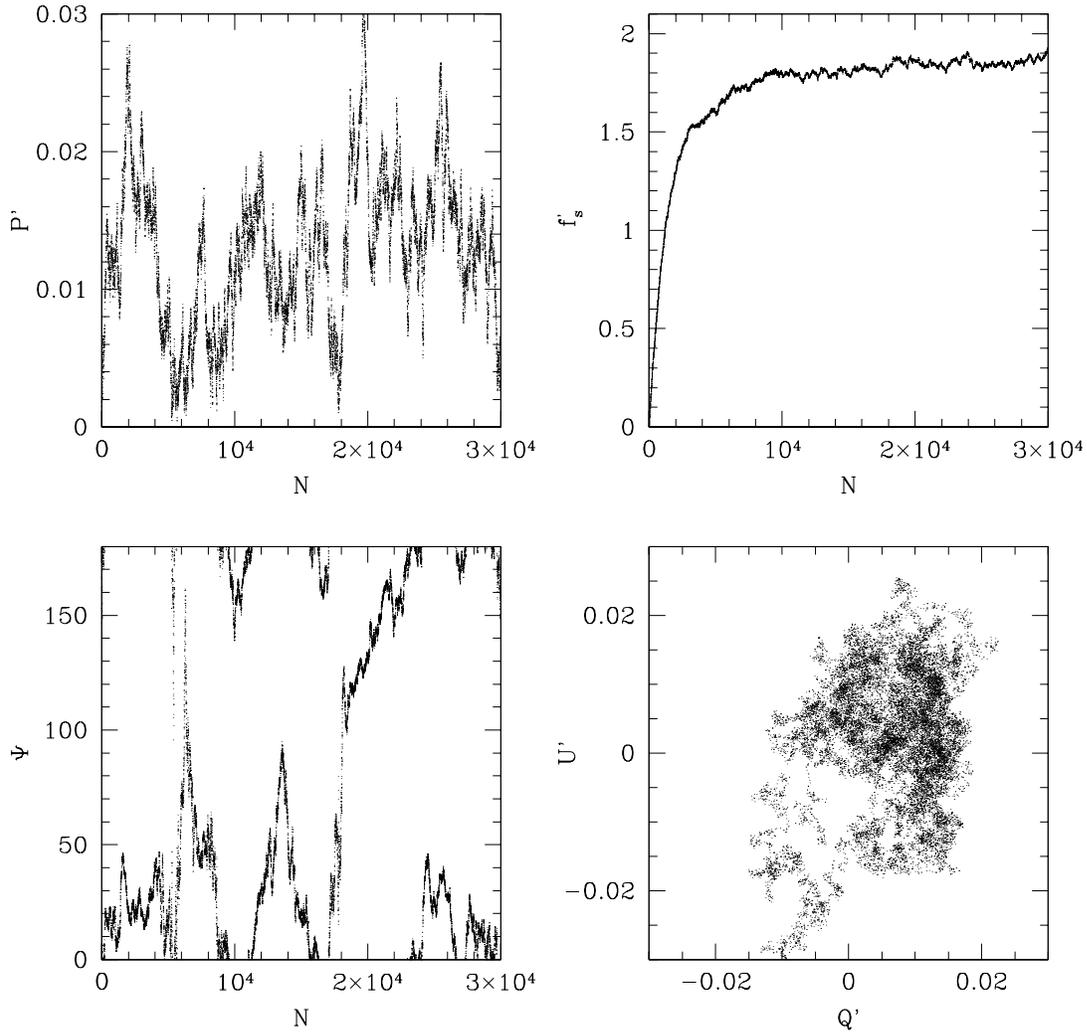}
\caption{The figures show respectively, instant by instant, model results vs number $N$ of clumps emitted thus far for the  following: upper left panel -- polarization $p'=p/p_0$;  upper right -- scattered light $f_{\rm s}^{'}=f_{\rm s}/f_0$;  lower left panel -- polarization position angle $\psi$;  lower right panel -- $Q'=Q(N)/p_0$ versus $U'=U(N)/p_0$. $\beta=0.5$ and $\nf=1000$ are applied.}
   \label{fig6}
\end{figure}

   \begin{figure}
\plotone{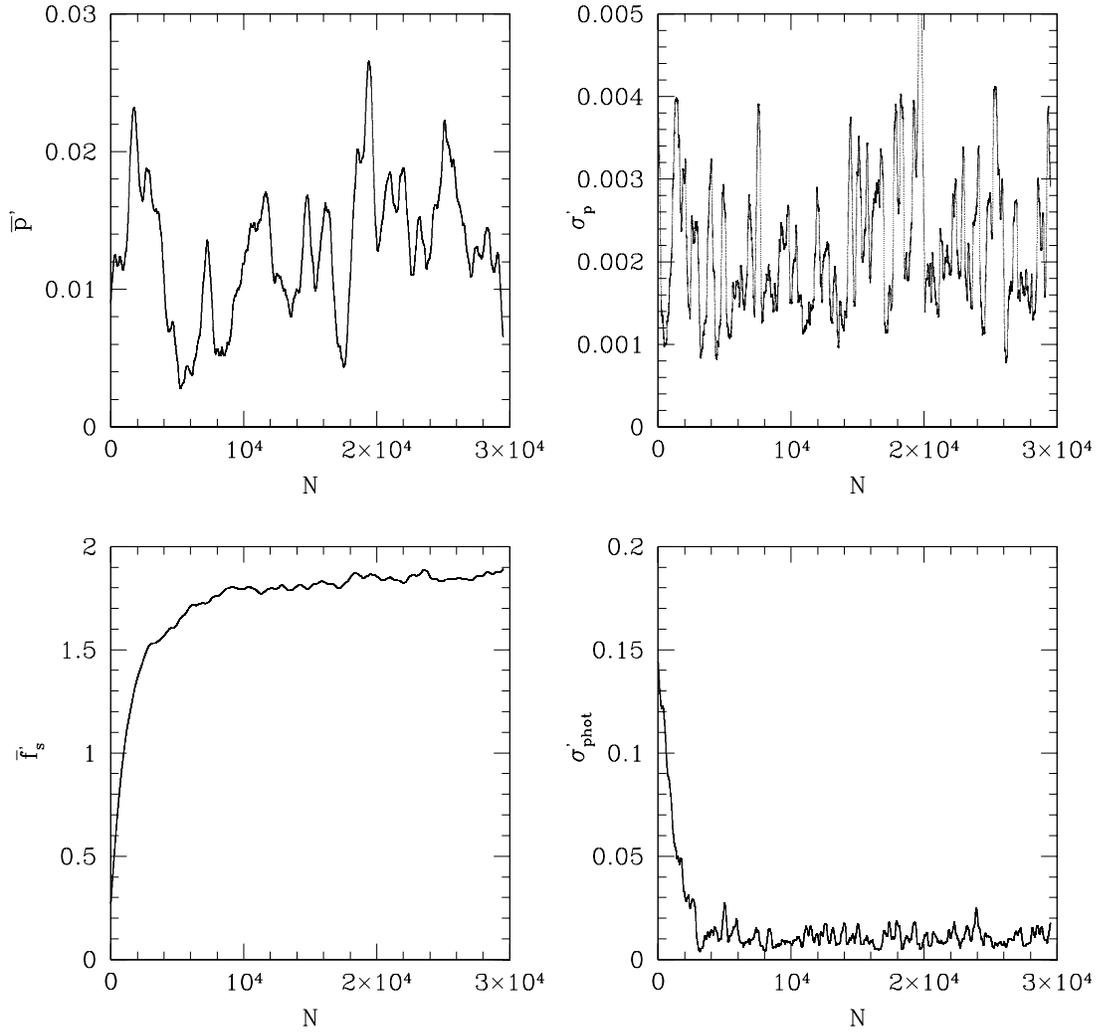}
\caption{ Based on the same data as Fig.~\ref{fig6}, here we show smoothed results versus number of clumps $N$ (increasing with time)
  for the following observables, with parameters as in Fig.~\ref{fig6}: (a) mean polarization $\bar p'=\bar p/p_0$; (b) variance
  of polarization $\sigma_{\rm p}^{'}=\sigma_{\rm p}/p_0$; (c) mean scattered light fraction $\bar {f_{\rm s}}^{'}=\bar {f_{\rm s}}/f_0$; (d) variance of scattered 
light $\sigma_{\rm phot}^{'}=\sigma_{\rm phot}/f_0$.}
   \label{fig7}
\end{figure}

   \begin{figure}
\plotone{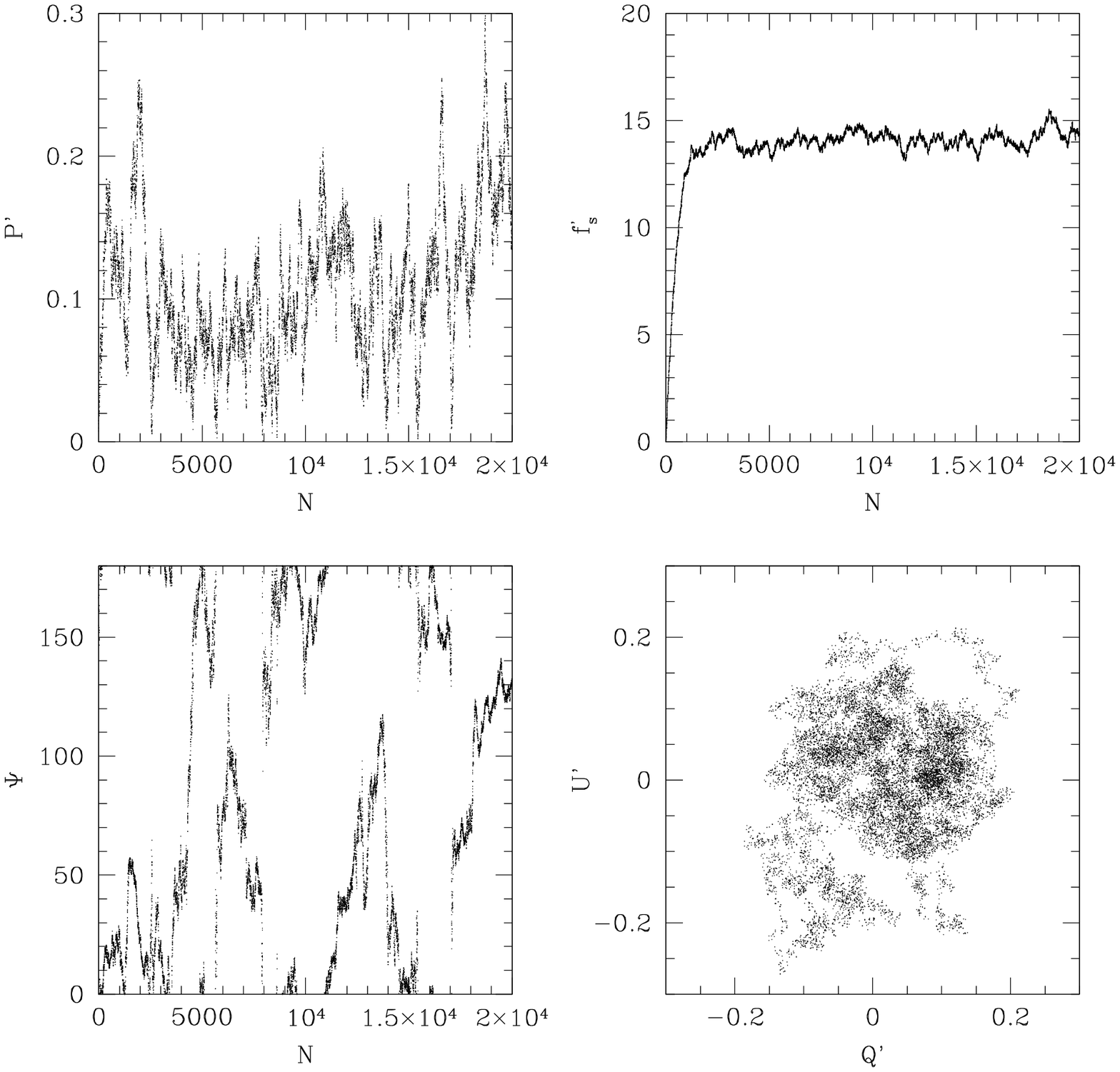}
\caption{The figures are similar to Fig.~\ref{fig6}, except $\beta=1$ and $\nf=200$ are applied.}
   \label{fig8}
\end{figure}

   \begin{figure}
\plotone{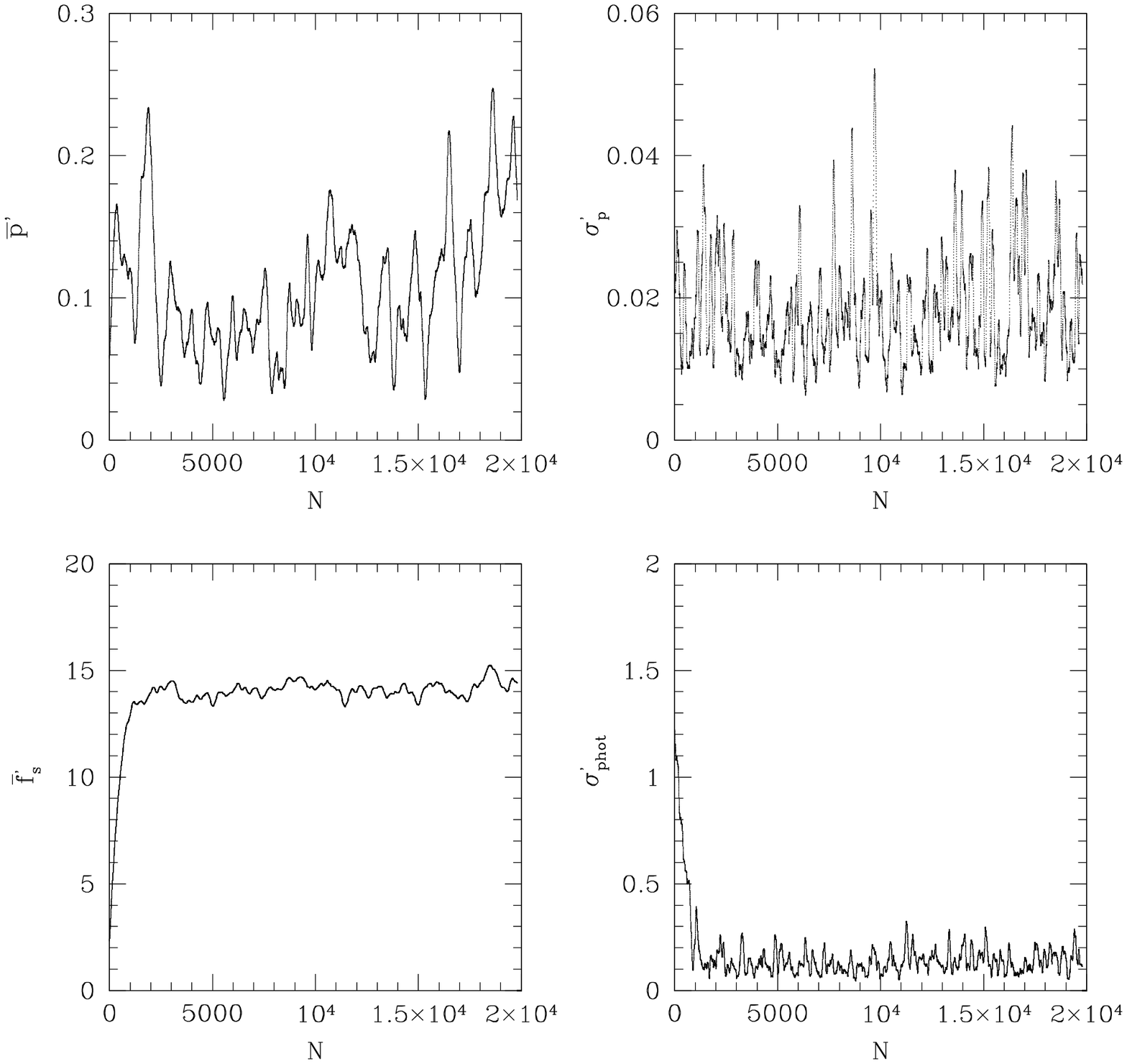}
\caption{ The figures are similar to Fig.~\ref{fig7}, except $\beta=1$ and $\nf=200$ are applied.}
   \label{fig9}
\end{figure}

   \begin{figure}
\plotone{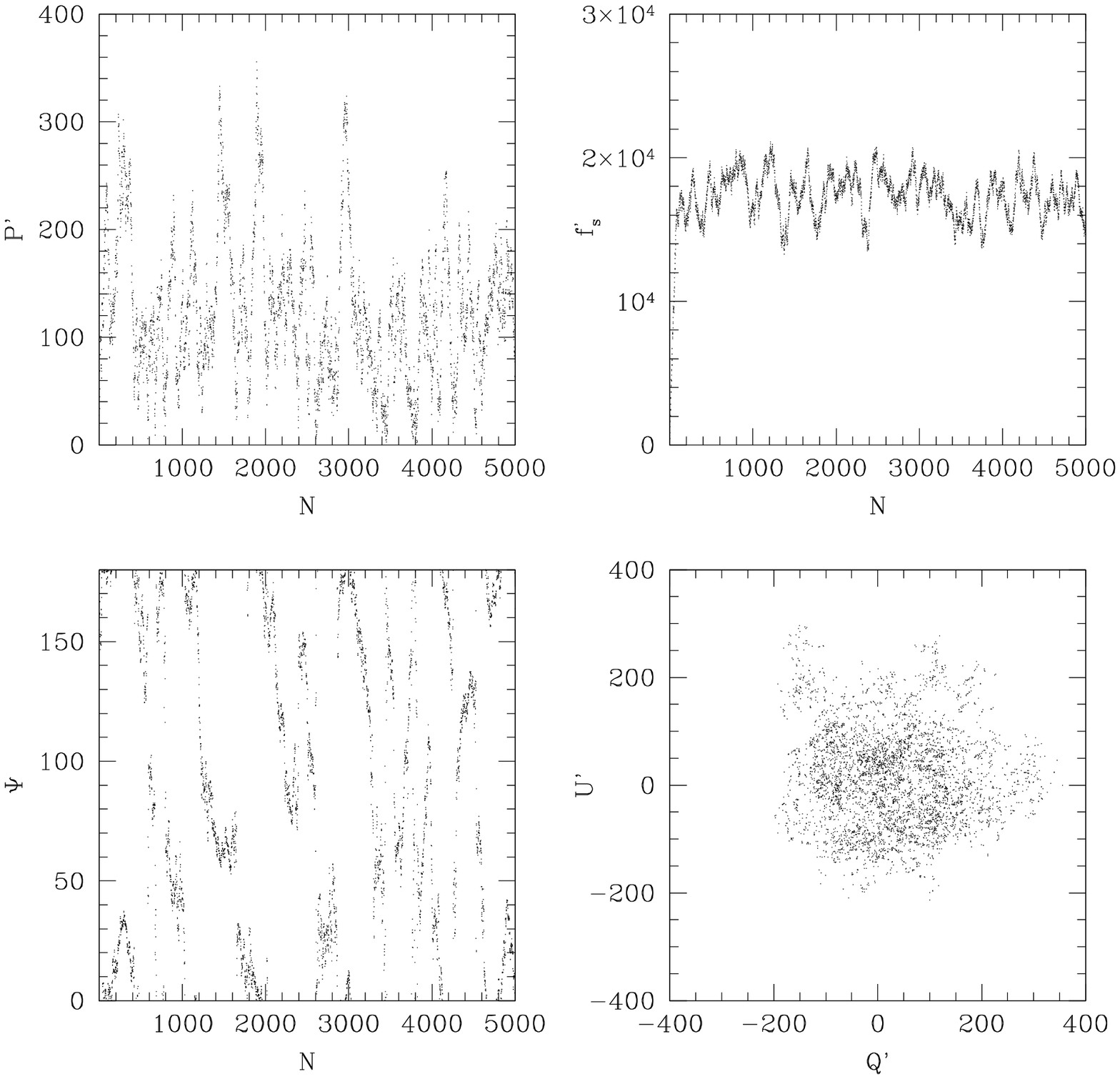}
\caption{The figures are similar to Fig.~\ref{fig6}, except $\beta=2$ and $\nf=1$ are applied.}
   \label{fig10}
\end{figure}

   \begin{figure}
\plotone{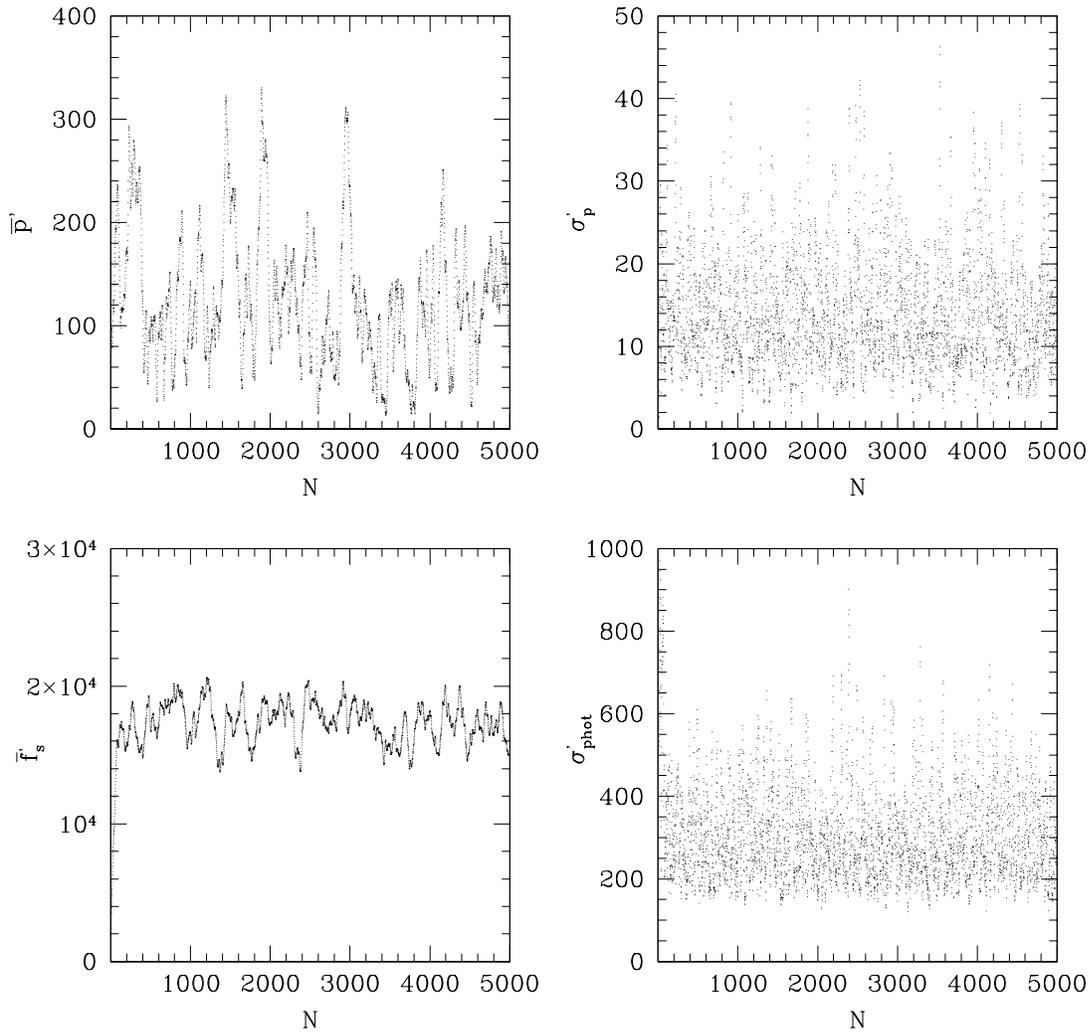}
\caption{ The resulting figures are similar to Fig.~\ref{fig7}, except $\beta=2$ and $\nf=1$ are applied. }
   \label{fig11}
\end{figure}

   \begin{figure}
\plotone{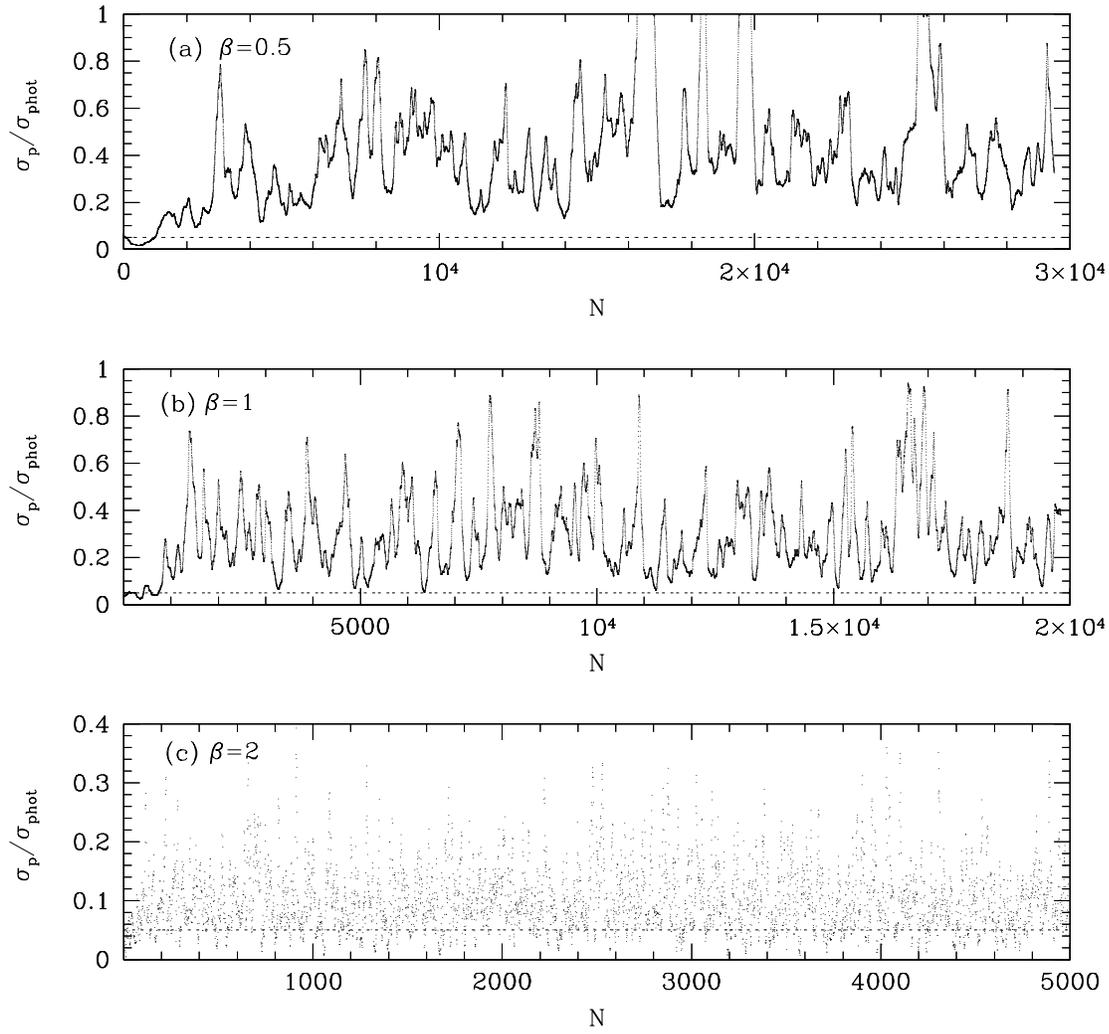}
\caption{ Ratio ($\sr=\sigma_{\rm p}/\sigma_{\rm phot}$) of polarimetric to
  photometric standard deviations vs number of clumps $N$. The upper panel (a) is for $\beta=0.5$ and $\nf=1000$. 
The middle panel (b) is for $\beta=1$ and $\nf=200$. The lower panel (c) is for $\beta=2$ and $\nf=1$. The steady 
mean value $\sr$ about 0.05 can be sustainable in case of $\beta=2$ and $\nf=1$. }
   \label{fig12}
\end{figure}

   \begin{figure}
\plotone{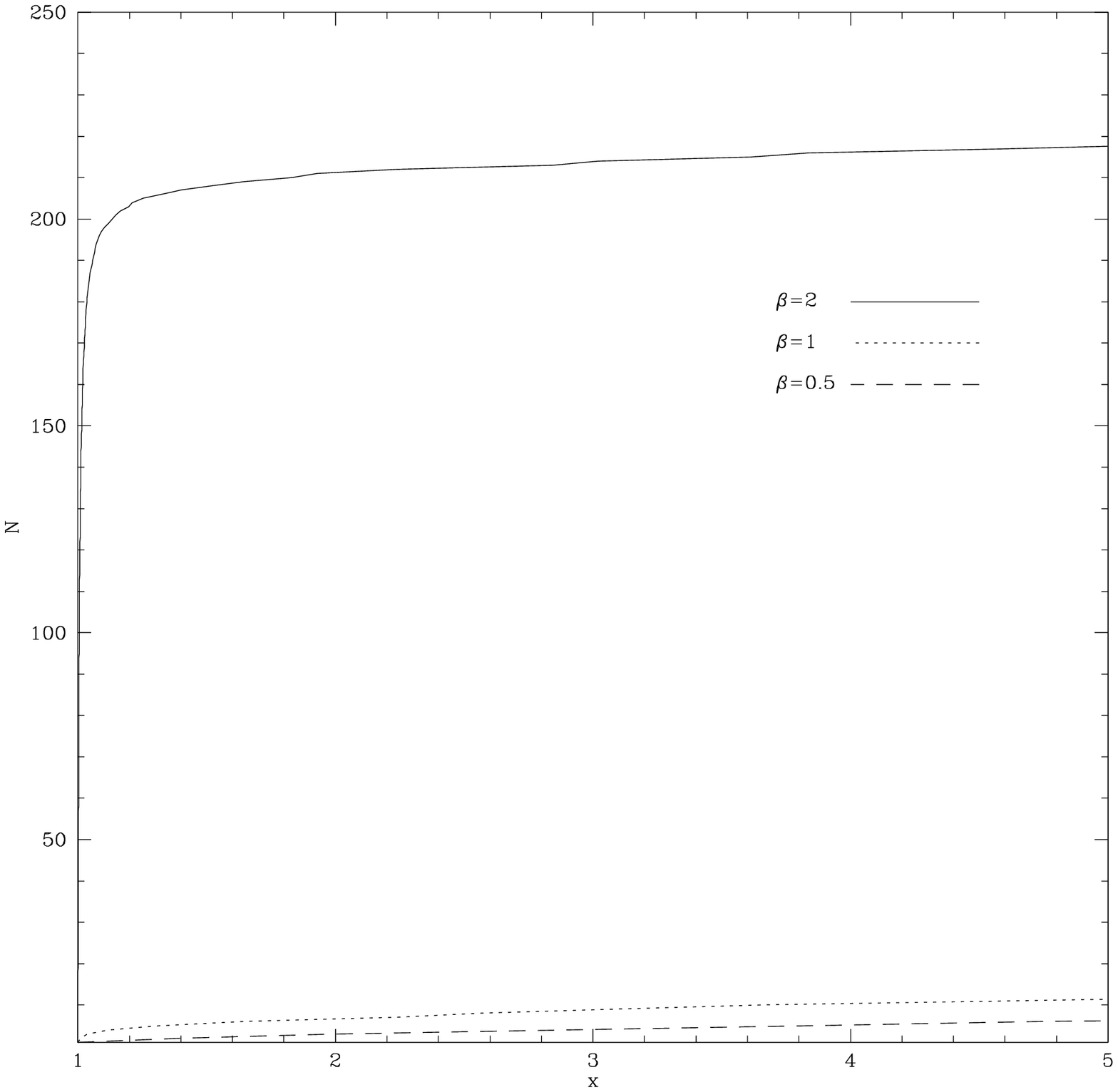}
\caption{ The number of clumps $N$ vs the radial distance ($x$) in case of $\nf=1$. The solid line is for $\beta=2$, the dotted line is for $\beta=1$, and the dashed line is for $\beta=0.5$. Due to the ``dwelling" time around the star is longer in case of larger $\beta$, the number of clumps are more, in comparison with the smaller $\beta$ case.}
   \label{fig13}
\end{figure}


\begin {thebibliography}{99}

\bibitem [1981]{abbo81}Abbott, D. C., Bieging, J. H., \& Churchwell, E. 1981, \apj, 250, 645

\bibitem [1994]{bro94} Brown, J.~C. 1994, in: Quebec Workshop on Instability
and Variability in Hot Star Winds. Eds. Moffat, A.F.J. and St-Louis, N., APSS, 221,357

\bibitem [1977]{bro77} Brown, J.~C. \& McLean, I.~S. 1977, \aap, 57, 141

  \bibitem [1989]{bro89} Brown, J.~C., Carlaw, V.~A., \& Cassinelli, J.~P. 1989, \aj 344, 341

  \bibitem [2000]{bro00} Brown, J.~C., Ignace, R., Cassinelli, J.~P. 2000, \aap, 356, 619

  \bibitem [1995]{bro95} Brown, J.~C., Richardson, L.~L., Antokhin, I., Robert, C., Moffat, A.~F.~J., St-Louis, N. 1995, \aap 295,725


  \bibitem [1998]{bro98} Brown, J.~C., Richardson, L.~L., Ignace, R.,
Cassinelli, J.~P. 1998, \aap, 330, 253

  \bibitem [2004]{bro04}Brown, J. C., Cassinelli, J. P., Li, Q., Kholtygin, A. F., \& Ignace, R. 2004, \aap, 426, 323


 \bibitem [1987]{Cas87} Cassinelli, J.~P., Nordsieck, K.~H., Murison, M.~A. 1987, \apj, 317, 290

  \bibitem [2008]{Cas08}Cassinelli, J. P., Ignace, R., Waldron, W.L., Cho, J., Murphy, N. A., \& Lazarian, A. 2008, \apj, 683, 1052

 \bibitem [2007]{dav07} Davies, B., Vink, J.S.,
\& Oudmaijer, R.D. 2007, \aap, 469, 1045

 \bibitem [2005]{des05} Dessart, L. \& Owocki, S.P. 2005, \aap, 432, 281

\bibitem [1987]{dri87} Drissen, L., St.-Louie, N., Moffat, A.~F.~J., Bastien, P. 1987, \apj, 322, 888

\bibitem [1992]{dri92} Drissen, L., Robert, C., Moffat, A.~F.~J. 1992, \apj, 386, 288

\bibitem [1998]{ham98} Hamann, W.R., \& Koesterke, L. 1998, \aap, 335, 1003

  \bibitem [2003]{ign03} Ignace, R., Quigley, M. F., \& Cassinelli, J. P. 2003, \apj, 596, 538

  \bibitem [1999]{lep99} Lepine, S., Moffat, A.~F.~J., 1999, \apj, 514, 909

  \bibitem [2000]{lep00}Lepine, S., Moffat, A.F.J., St-Louis, N., Marchenko, S.V.,   Dalton, M.J., Crowther, P.A., Smith, L.J., Willis, A.J., Antokhin, I.I., \& Tovmassian, G.H. 2000, \aj, 120, 3201

  \bibitem [2000]{li00}Li, Q., Brown, J. C., Ignace, R., Cassinelli, J. P., \& Oskinova, L. M. 2000, \aap, 357, 233

  \bibitem [1987]{lup87} Lupie, O.~L., Nordsieck, K.~H. 1987, \aj, 92, 214

  \bibitem [2006]{mar06} Marchenko, S. V., Moffat, A. F. J., St-Louis, N., \& Fullerton, A. W. 2006, \apj, 639, L75

  \bibitem [Moffat \& Robert(1992)]{mof92} Moffat, A.~F.~J, Robert, C. 1992, ASP \#22, 203	

\bibitem[2007]{osk07}Oskinova, L.M., Hamann, W.-R., \& Feldmeier, A. 2007, \aap, 476, 1331

 \bibitem [2006]{owo06} Owocki, S. P., Cohen, D. H. 2006, \apj, 648, 565


  \bibitem [1996]{ric96} Richardson,L.~L., Brown, J.~C., Simmons, J.~F.~L. 1996, \aap, 306, 519

  \bibitem [1992]{rob92} Robert, C. 1992, Ph.D. Thesis, Univ. de Montreal

  \bibitem [1989]{rob89} Robert, C., Moffat, A.~F.~J., Bastien, P., Drissen, L., St.-Louie, N. 1989, \apj, 347, 1034

  \bibitem [1987]{stl87} St.-Louis, N., Drissen, L., Moffat, A.~F.~J., Bastien, P., Tapia, S. 1987, \apj, 322, 870

  \bibitem [1991]{tay91} Taylor, M., Nordsieck, K.~H., Schulte-Ladbeck, R.~E., Bjorkman, K.~S. 1991, \aj, 102, 1187

  \end {thebibliography}
\end{document}